\documentclass[10pt,letterpaper]{article}
\usepackage[top=0.85in,left=2.75in,footskip=0.75in]{geometry}

\usepackage{amsmath,amssymb}

\usepackage{changepage}

\usepackage[utf8x]{inputenc}

\usepackage{textcomp,marvosym}

\usepackage{cite}

\usepackage{nameref,hyperref}

\usepackage{microtype}
\DisableLigatures[f]{encoding = *, family = * }

\usepackage[table]{xcolor}

\usepackage{array}

\newcolumntype{+}{!{\vrule width 2pt}}

\newlength\savedwidth

\newcommand\thickhline{\noalign{\global\savedwidth\arrayrulewidth\global\arrayrulewidth 2pt}%
\hline
\noalign{\global\arrayrulewidth\savedwidth}}

\raggedright
\usepackage[aboveskip=1pt,labelfont=bf,labelsep=period,justification=raggedright,singlelinecheck=off]{caption}

\makeatletter
\renewcommand{\@biblabel}[1]{\quad#1.}
\makeatother

\usepackage{lastpage,fancyhdr,graphicx}
\usepackage{epstopdf}
\fancyheadoffset[L]{2.25in}
\fancyfootoffset[L]{2.25in}
\begin{document}
\vspace*{0.2in}

\begin{flushleft}
{\Large
\textbf\newline{A physics-based model of swarming jellyfish} 
}
\newline
\\
Erik Gengel \textsuperscript{1},
Zafrir Kuplik \textsuperscript{2,3},
Dror Angel \textsuperscript{3},
Eyal Heifetz \textsuperscript{1}
\\
\bigskip
\textbf{1} Department of Geophysics, Porter School of the Environment and Earth Sciences, Tel Aviv University, Tel Aviv 69978, Israel.
\\
\textbf{2} The Steinhardt Museum of Natural History, Tel Aviv University, 12 Klausner Street, Tel Aviv
\\
\textbf{3} The Leon Recanati Institute for Maritime Studies, University of Haifa, Mount Carmel, Haifa 3498838, Israel
\\
\bigskip

%
%





* egiu@gmx.de

\end{flushleft}
\section*{Abstract}
We propose a model for the structure formation of jellyfish swimming based on active Brownian particles. We address the phenomena of counter-current swimming, avoidance of turbulent flow regions and foraging. We motivate corresponding mechanisms from observations of jellyfish swarming reported in the literature and incorporate them into the generic modelling framework. The model characteristics is tested in three paradigmatic flow environments.



\section*{Introduction}

Scyphozoans populate the oceans since the late proterozoic eon and according to geological records they are one of the first multicellular organisms on planet Earth \cite{van2014origin}. {The life cycle of scyphozoans is complex including in many species the tiny, cryptic, benthic stage of the polyp. But, most prominent is the planktonic stage of the medusa}, better known as \textit{jellyfish} \cite{helm2018evolution, lotan1992life}. Despite their appearance, medusae are among the most efficient swimmers in the oceans \cite{costello2021hydrodynamics,gemmell2013passive} and their massive occurrence in different habitats has major impacts on the biosphere, tourism, and economics \cite{angel2016local,schrope2012attack,streftaris2006alien,nakar2011economic}. While jellyfish have existed throughout human civilization, the adverse effects of climate change on the ocean habitats are hypothesized to intensify jellyfish blooms and the associated impacts \cite{attrill2007climate, angel2016local}.

The fact that {scyphomedusa} establish swarms and interact with humans on numerous levels has motivated researchers to investigate the relationships between large jellyfish aggregations and environmental conditions \cite{zhang2012associations, baliarsingh2020review, heim2019salinity,edelist2020phenological, houghton2006developing, cimino2018jellyfish}. Results of these studies have mainly contributed to empiric models for prediction of jellyfish proliferation in estuaries and bays \cite{brown2002forecasting, ruiz2012model, fossette2015current} or along open coasts \cite{aouititen2019predicting, prieto2015portuguese, nordstrom2019tracking}. These approaches represent the coarse end of a spatio-temporal continuum of models, ranging from regional scale to the level of single-agent swimming. Indeed, many studies have examined the efficiency of different propulsion mechanisms in single medusae, experimentally \cite{malul2019levantine, dabiri2005flow, gemmell2015control} and computationally \cite{hoover2017quantifying,yuan2014numerical, wilson2009lagrangian, hoover2019pump,dular2009numerical,sahin2009arbitrary, hoover2015numerical,park2015dynamics}, based on detailed observations and direct numerical simulations of bell oscillation and the resulting vortex dynamics in the surrounding water. 

In addition, jellyfish respond to a variety of environmental stimuli, affecting the emergence and movement of swarms as a whole. Observations suggest that jellyfish modify their swimming behavior with respect to: the location of other individuals \cite{hamner2009review}, the water temperature \cite{ruiz2012model,heim2019salinity}, salinity \cite{edelist2020phenological,zhang2012associations}, advection by water currents, turbulence  \cite{malul2019levantine, fossette2015current} and the presence of prey \cite{arai1991attraction,matanoski2001characterizing}. However, there is a large mismatch, between what is known regarding jellyfish ecology and the accuracy of jellyfish swarm prediction based on circulation and distribution models. Thus, there is a great need for a theoretical modelling framework that ties together single-agent dynamics, response and swarm behavior, applicable in large scale circulation models \cite{lotan1992life,omori2004taxonomic, nooteboom2020resolution, bryan1995midlatitude, delworth2012simulated}. 

Thus, in this paper, we {pursue two research goals: First, we propose a mechanistic model framework for the swarming dynamics of jellyfish based on active Brownian particles (ABPs) \cite{schweitzer2003brownian,romanczuk2012active}.} A key aspect of the model is that {it reaches far beyond existing approaches for modelling of jellyfish proliferation (see Tab.~\ref{table0} below): Jellyfish in the model are considered to move actively and they can make decisions based on environmental information. In particular, we pursue the idea that the environmental stimuli are processed by the neuronal network of a jellyfish. This network generates a coherent strong pulsation rhythm which gives rise to the bell movement in many jellyfish species. Accordingly, we consider a jellyfish in our model to be a moving nonlinear oscillator, or \textit{swarmalators} \cite{o2017oscillators,o2019review,hong2021coupling}. A swarmalator is the next step towards modelling of agent-based swarm phenomena, combining active-matter research and the theory of nonlinear oscillators \cite{pikovsky2001synchronization,pikovsky2021transition, uriu2013dynamics,zheng2021transition}. In particular the latter allows us to incorporate several important biological features into the model, for example the species-dependent response to external stimuli.}

{The second and main research goal is to adapt the introduced parameters of the model according to the existing literature and experience from field observations \cite{malul2019levantine,arai1991attraction,zafrirpersonal}. Here we integrated data collected for species of the orders Rhizostomeae and Semaeostomeae which are known for their swarming behavior but differences in the bell shape (prolate or oblate) and swimming patterns (jet-propulsion or rowing-propulsion) in adult medusae.}

{On the one hand, the particle-based description allows to extract a consistent mathematical description of individual behavior by means of data analysis techniques for oscillatory systems \cite{rosenblum2001phase,cestnikinfering2020, kralemann2007uncovering, gengel2021phase, smeal2010phase,Rosenblum-Pikovsky-01}. On the other hand, the ensemble dynamics can be tested in a transient development process of prediction and continuous parameter adjustment, using physics-informed machine learning \cite{kashinath2021physics,mcilwaine2021jellynet, martin2020jellytoring, albajes2011jellyfish} where the parameter set of this work serves as a starting point. }
{Given the fact that the fields of active matter dynamics and jellyfish ecology have been largely disconnected up to now, a certain amount of intuition is needed to construct an overarching model and to design numerical tests for its validation. As a consequence,} we exploit only a minimal set of mechanisms, related to environmental inputs, while we leave out most of the agent-agent interactions. {Nevertheless, the resulting active oscillatory swimmers are able to mimic biological phenomena found in swimming jellyfish \cite{albert2011s}.} Our analysis {and characterization of the model performance} focuses on theoretical and computational strategies to investigate pattern formation in networks of moving oscillators from the perspective of active-matter research \cite{vicsek1995novel,wolgemuth2008collective, aranson2013active, cavagna2014bird, canizo2010collective, ramaswamystatactive2010, stenhammar2017role, grossmann2014vortex, reinken2019anisotropic} {and tries to relate these outcomes to the biological background. We elaborate in the end of the text which theoretic quantities should be extracted from experimental data to strengthen the observational evidence for active-jellyfish modelling.}

\section*{Methods}

We obtain the flow field $U(\mathbf{x},t)$ (hereafter $\mathbf{x}$ denotes position on the horizontal plane and $t$ denotes time) from the incompressible Navier-Stokes equations which we simulate by means of a second-order in time and space method on a staggered grid using a Successive Over-Relaxation (SOR) pressure-solver \cite{kampanis2006staggered,kundufluiddyn}. The dynamics of the prey concentration $F(\mathbf{x},t)$ is modelled by an advection-diffusion equation coupled to the flow. We fix the diffusion coefficient of the prey to be $D_F=0.001$\ m$^2$\ s$^{-1}$ \cite{okubo1971oceanic}.

The trajectories of the jellyfish are simulated by the second-order stochastic Heun-method \cite{mannella2002integration}. We couple the dynamics of the jellyfish to the flow by using a bi-linear interpolation of the local field quantities onto the position of each single jellyfish agent $j$ \cite{press1992numerical} so that for instance $F_j=F(\mathbf{x}_j,t)$. 

For the statistical analysis of the swarming behavior we make use of ensemble averages, the Pearson correlation \cite{adler2010quantifying} and the indication number:
\begin{equation}
 \begin{aligned}
    \overline{W}_M &= \frac{1}{M}\sum_{j=0}^{M-1} W_j, \\ \text{Corr}_{W^1,W^2} &= \left( \overline{W^1W^2}_N - \overline{W^1}_N\ \overline{W^2}_N \right)/(\sigma^1\sigma^2), \\
    \mathcal{N}(W) &= \overline{\mathcal{H}(W-\hat{W})}_N \; . 
     \end{aligned}\label{eq: pearson}
\end{equation}
Here $j$ denotes the serial number of a jellyfish agent and $M\leq N$ is the overall number of agents considered for averaging. {$W$ can be any simulated field variable, e.g.\ the concentration of prey.} $N$ is the total number of agents. In reality jellyfish swarms are composed of hundreds of thousands of individuals in the open sea, or just several few individuals in controlled tank experiments \cite{mackie1981swimming, nath2017jellyfish, gemmell2015control, baliarsingh2020review}. In this work we use $N=128$ agents as a tradeoff between theoretical demands and experimental limitations. 

To obtain statistically significant results, we average {values of Eq.~\eqref{eq: pearson}} over $K=16$ model runs using a Gaussian kernel with a window of $15$\ seconds \cite{ghosh2018kernel}. The resulting double average is denoted as $\langle W \rangle \equiv \langle \overline{W}_M \rangle_K$. $\sigma^{1}$ and $\sigma^{2}$ are the standard deviations of variables $W^1$ and $W^2$. {${\cal N}(W)$ measures the fraction of jellyfish that can be found in domains fulfilling the condition $W>\hat{W}$. To compare the performances of active jellyfish and passive tracers, we use the ensemble average for passive tracers $\hat{W} \equiv \overline{W}_{\text{passive},N}$ if not stated otherwise. For example regarding prey searching, we expect that jellyfish will try to maximize their access to ambient prey while passive tracers will not show any response. Thus, when for a majority of jellyfish $F_j>\hat{F}$ holds, ${\cal N}(F) \approx 1$ indicates an excess of preying performance and most jellyfish are situated in regions where the prey concentration is at least higher than $\hat{F}$. An exemplary separation of the flow domain is depicted in Fig.~\ref{fig: vortex avoidance}. In technical terms, the separation is represented by the Heaviside function $\mathcal{H}$ which returns unity if $W>\hat{W}$ and zero otherwise.}

\subsection*{Towards a description of active swarming jellyfish} \label{sec model}

We model a single jellyfish agent as an active over-damped particle at horizontal position $\mathbf{x} = (x,y)^{\top}$ and time $t$, having a velocity $\mathbf{v}(\mathbf{x},t)$ and orientation $\theta(\mathbf{x},t)$ \cite{schweitzer2003brownian, romanczuk2012active}. In the following, we propose a dynamic model for the position and the orientation of the agent which allows us to capture paradigmatic behaviors of jellyfish. 

Jellyfish process environmental information by a neuronal network of several thousand neurons. The dynamics of this network can be regarded as a perturbed internal dynamics \cite{pallasdies2019single, satterlie2011jellyfish, garm2006rhopalia, hoover2021neuromech} in which neurons fire coherently and act as a single large oscillator to drive the bell pulsation. Moreover, it has been shown that the neuronal network of jellyfish features properties of a circadian oscillator \cite{nath2017jellyfish, Winfree-80}. Thus, the bell pulsation can be regarded as the resulting average network oscillation \cite{malul2019levantine, gemmell2013passive, pikovsky2001synchronization}. For such a process it has been shown theoretically \cite{watanabe1994constants,kuramoto2003chemical, wilson2016isostable, hansel1993phase,levnajic2010phase,hagos2019synchronization,pikovsky2001synchronization} and empirically \cite{topccu2018disentangling, Kralemann_etal-13} that a low-dimensional description, in terms of a phase variable $\varphi(t)$ is possible and beneficial for a swift but reliable model development. Moreover, phase dynamics models are computationally light, but at the same time a reliable simplification, as they are based on observational evidence and a fit to the theory \cite{hansel1993phase, gengel2021phase, rosenblum2001phase,Kralemann_etal-13}. 

{Another aspect of jellyfish motion is that swimming patterns can change abruptly} in response to stimuli \cite{albert2011s,zafrirpersonal}. In such cases jellyfish will switch from a state of relative inactivity into a state where the frequency and the strength of the bell oscillation increase and the swimming trajectories encompass a significantly larger volume of fluid \cite{bailey1983laboratory, matanoski2001characterizing, hays2012high}. Such changes in swimming patterns can even cause a swarm to cross flow barriers \cite{arai1991attraction}. {Following such periods of elevated activity, jellyfish will return to a state of relative quiescence.} To capture this behavior, we introduce an additional activity variable ${\cal A}$ which causes parametric switching \cite{hagos2019synchronization, hodgkin1952quantitative, morris1981voltage}. In accordance with theoretical studies on relaxation oscillators, this variable can be regarded as the leading-order amplitude perturbation from a limit cycle \cite{mauroy2013isostables,gengel2020high,wilson2016isostable}. {We assume that this degree of freedom in the jellyfish dynamics is stimulated by external inputs and that it decays exponentially at a rate $\lambda_{\cal A}$, in order to ensure it returns to a quiescent state.} 

\begin{figure}[!h!]
\centering
\includegraphics[width=0.99\columnwidth, angle=0]{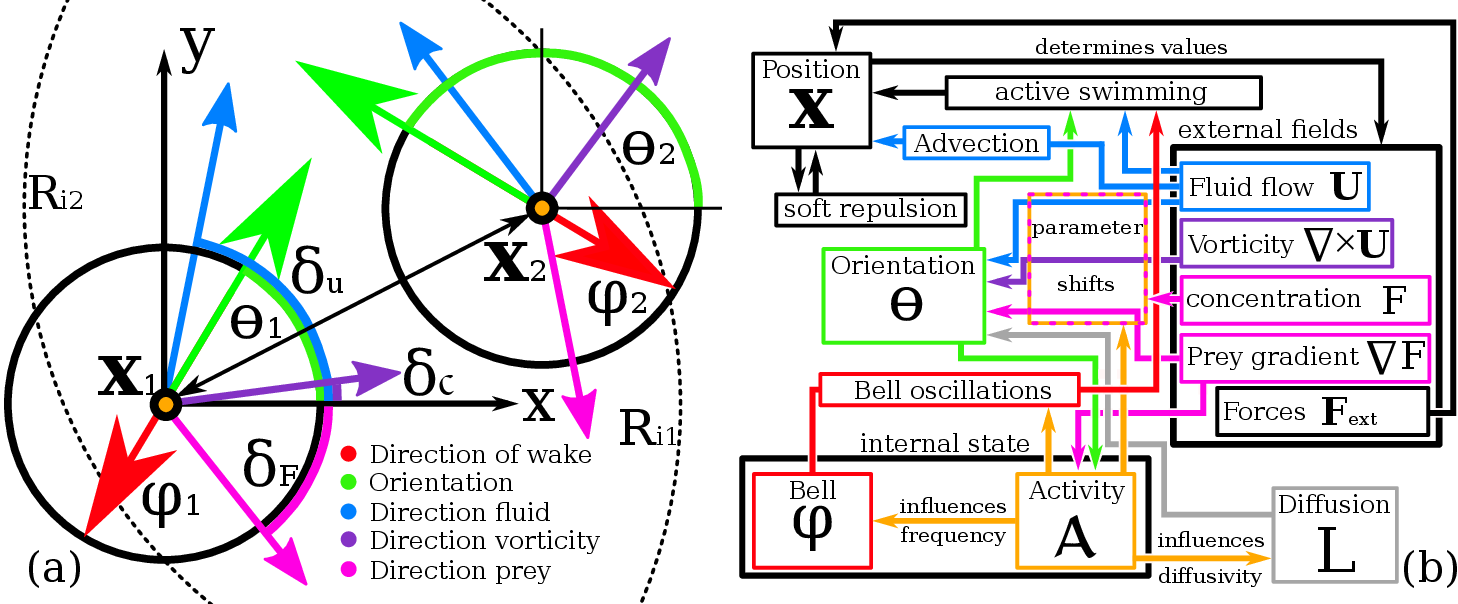}
\caption{Scheme of the jellyfish agent model. Panel (a): A pair of two agents. Dashed lines indicate the interaction radius $R_i$, The arrow connecting both agents indicates the soft-core repulsion. Colored arrows indicate different orientations according to panel (b). Panel (b): Flow diagram of the state variables $\mathbf{x}$, $\varphi$, $\theta$ and activity ${\cal A}$. Included model mechanisms are advection, swimming against the flow, soft-core repulsion, avoidance of walls, bell oscillations, avoidance of turbulence, local random motion and behavioral shifts due to activity changes (presence of prey). External drivers of the model are the fluid velocity $\mathbf{U}_j$, the absolute vorticity $|{\cal C}|_j$ the gradient of prey $\nabla F_j$ and the prey concentration $F_j$.}
\label{fig schematic model}
\end{figure}

Following these generic considerations, {the dynamics of a single }jellyfish $j$ in a swarm of $N$ agents is {described by} the four differential equations:  
\begin{equation}
\begin{aligned}
\dot{\mathbf{x}}_j & = \mathbf{v}_j(\mathbf{X},\boldsymbol{\theta}, \boldsymbol{\varphi}, {\cal A}_j, F_j,\mathbf{U}_j,t) \\
\dot{\theta}_j &= {\cal G}_j(\mathbf{X},\boldsymbol{\theta},{\cal A}_j,F_j,|\mathcal{C}|_j,t) \\
\dot{\varphi}_j &= \omega_j({\cal A}_j) + H_j(\mathbf{X},\boldsymbol{\theta},\boldsymbol{\varphi},\mathbf{\cal A}_j) \\
\dot{\cal A}_j &= -\lambda_{\cal A} {\cal A}_j + I_j({\theta},F_j)
\end{aligned} \; . \label{eq swarming jelly}
\end{equation}
Here, we have used the ensemble notation $\mathbf{X} = [\mathbf{x}_1,\ldots,\mathbf{x}_N]$, $\boldsymbol{\varphi} = [\varphi_1,\ldots,\varphi_N]$, $\boldsymbol{\theta} = [\theta_1,\ldots,\theta_N]$. 
$F$ indicates the prey concentration and $\mathbf{U}$ denotes the water horizontal current vector. $\mathcal{C} = \nabla \times \mathbf{U}$ is the vorticity of the water flow. We take its absolute value as a simplified measure for the amount of turbulence in the flow \cite{buaria2020vortex}. In the following we exemplify the generic set of Eqs.~\eqref{eq swarming jelly}. For convenience, a scheme of this jellyfish agent model is illustrated in Fig.~\ref{fig schematic model}. {A simplistic comparison of the model to already existing research on active matter and jellyfish dynamics is presented in Tab.~\ref{table0}.} Next we explain the bottom-up rational of the model. 

\begin{table}[!ht!]
\begin{adjustwidth}{-0cm}{0cm} 
\centering
\caption{
{\bf Comparison of the swarming model Eqs.~(\ref{eq activity state}-\ref{eq: orientation state}) and Eqs.~(\ref{eq: responses eps}-\ref{eq attr rep}) to previous simulation approaches}}\label{tab: params inputs}
\begin{tabular}{|l|l|l|l|}
\hline
{\bf } & {\bf Model } & {\bf Jellyfish} & {\bf Active matter}\\ \thickhline
 Type & ABPs & Passive tracers \cite{prieto2015portuguese,el2020modelling} & ABPs \cite{schweitzer2003brownian} \\
  & Swarmalators \cite{o2017oscillators} & Average models \cite{brown2002forecasting,ruiz2012model} & Continuum models  \\
  &  & Empiric models \cite{albajes2011jellyfish} & \cite{tonertu,wolgemuth2008collective} \\
  & & Single-agent models \cite{hoover2017quantifying} & \\
\hline
 Dynamic & Position $\mathbf{x}_j$ & Position & Position, velocity \\
 variables & Orientation $\theta_j$ & Concentration & Density \cite{tevrugtpredict23} \\
 & Phase $\varphi_j$ & Jellyfish geometry \cite{yuan2014numerical} & Polarization \cite{cavagna2014bird} \\
 & Activity ${\cal A}_j$ &  & \\
 \hline
 Interactions & Parameterized & -- & Parameterized \cite{canizo2010collective} \\
 &  &  & Hydrodynamic \cite{negro2022inertial} \\
 \hline
 Forcings & Flow $\mathbf{U}_j$ & Flow $\mathbf{U}_j$ \cite{prieto2015portuguese,fossette2015current} & Flow $\mathbf{U}_j$ \\
 & Vorticity ${\cal C}_j$ & Temperature \cite{ruiz2012model} & Temperature, light \\
 & Prey $F_j$ & & chemicals \cite{januswalt08} \\
 \hline
 Parameters & Dynamic in & Constant & Constant or  \\
 & response to &  & forced \cite{petkoski2012kuramoto,childs2008stability} \\
 & environment & & \\
 \hline
 Scales & $1$\ s -- $1$\ h & $1$\ s -- $1$\ month  & $1$\ ms -- $1$\ h  \\
 & $10$\ cm -- $10$\ m & $10$\ cm -- $100$\ km & $1$\ nm -- $1$\ km \\
  \hline 
 Application & Understanding & Prediction of blooms in: & Pattern formation in \\ 
 & the interaction of & Estuaries, lagoons & swarms of: \\
 & jellyfish in tanks & bays \cite{brown2002forecasting,ruiz2012model}  & Birds, fish, locust, \\
 & (Local predictions) & Beaching events \cite{baliarsingh2020review, prieto2015portuguese} & fireflies \cite{locustariel15, fireermen91} \\
 & (Upscaling) & Hydrodynamics & Pedestrians \cite{helbing1995social} \\
 & & of swimming \cite{dabiri2005flow,hoover2015numerical} & Bacteria \cite{drescher2011fluid} \\
 \hline
\end{tabular}
\vspace{0.1cm}
\begin{flushleft}
 {Comparison between the proposed model, existing models for jellyfish and active matter models. In case of active matter, a focus is mostly put on biological applications. Scales represent rough estimates and refer to the mentioned studies. For the proposed model, we mention further steps in the development in brackets.}   
\end{flushleft}
\label{table0}
\end{adjustwidth}
\end{table}

\subsubsection*{Agent activity ${\cal A}$}

Here we take into account only the presence of prey as an incentive to activate the jellyfish dynamics and assume that the activity evolves according to: 
\begin{equation}
\dot{\cal A}_j = {-\lambda_{\cal A} {\cal A}_j + I_j({\theta},F_j)=} -\lambda_{\cal A} {\cal A}_j + |\nabla F_j|\frac{1}{2}\Big[|\cos(\theta_j-\delta_F)| + \cos(\theta_j-\delta_F)\Big]\; . \label{eq activity state}
\end{equation}
The decay parameter $\lambda_{\cal A}$ corresponds to the time for which an agent is active. We fix this value to $\lambda_{\cal A}=0.005$\ s$^{-1}$, {equivalent to a characteristic activity time} of roughly $3$\ minutes \cite{titelman2006feeding, zafrirpersonal}.
We assume that the excitation is a linear function of the magnitude of the local prey concentration gradient $|\nabla F_j|$. Additionally, the strength of the linear forcing is modulated by the relative orientation of an agent with respect to the direction of the prey concentration gradient. Defining the angle of that gradient as $\delta_F$ (satisfying $\tan(\delta_F) = {\partial_y F}/{\partial_x F}$), the term in the square bracket of Eq.~\eqref{eq activity state} ensures that only swimming towards the prey ($|\theta_j-\delta_F|<\pi/2$) evokes an increase in activity. We choose this approach to mimic a basic type of energy saving strategy, by favoring directions towards prey over others.

\subsubsection*{Bell oscillation $\varphi$}

The oscillatory movement of the bell is modelled by a phase oscillator \cite{hong2021coupling,o2019review,uriu2013dynamics,petkoski2012kuramoto,rodrigues2016kuramoto}
{
\begin{equation}
\dot{\varphi}_j = \omega_j({\cal A}_j) + H_j(\mathbf{X},\boldsymbol{\theta},\boldsymbol{\varphi},{\cal A}_j)\end{equation}
with a frequency $\omega_j({\cal A}_j)$. Additionally, the oscillator reacts to external perturbations according to a phase coupling function $H_j(.)$.} 

We allow the natural frequency of the bell oscillations to vary according to the activity level ${\cal A}$. In order to mimic this coupling, we use the response function:
\begin{equation}
 \mathcal{R}(a,b,S) = \frac{aS}{b+S} \label{eq: response}\; .
\end{equation}
{Here, $a$ and $b$ are the generic parameters of the response to be set later, and $S$ is a stimulus that evokes the response. Generally, ${\cal R}$ resembles a nonlinear activation function, commonly associated with dynamics in large neuronal networks and employed in machine learning \cite{karlik2011performance}. The reason for this widespread use is that $\mathcal{R}$ mimics a quasi-linear response at small stimuli (${\cal R} \sim {a\over b} S$) and a saturation behavior at large stimuli (${\cal R}\sim a$), typical for physiological limitations in biological systems (see Fig.\ref{fig: couplings} panel (a)).} The resulting natural frequency response is defined as:
\begin{equation} 
\qquad \omega_j = \omega_{j,0}[1+\mathcal{R}(f_{\varphi},f_{\varphi},{\cal A}_j)] \; , \label{eq: internal state}
\end{equation}
where $\omega_{j,0}$ is the natural frequency of an agent in an inactive state. In this study we fix $\omega_{j,0}=\omega_0=1.2 \text{ rad\ s}^{-1}$ \cite{malul2019levantine,titelman2006feeding} for all agents. Since the bell oscillation frequency in several species can depend on the size of jellyfish, future experimental work is needed to determine the evolution of its distribution for different species during different seasons \cite{el2020modelling}. Moreover, we seek to allow for a maximal physiological frequency response which we expect to observe when jellyfish escapes its predators. Thus, we fix the frequency response parameter to $f_{\varphi}=0.75$ \cite{hansson1995behavioural}. 
We use a single type of response function, Eq.~\eqref{eq: response}, throughout this study both, due to the lack of experimental data and in order to simplify our model. We expect however, that based on future observations, a variety of response functions will be required, similar to paradigmatic models of computational neuroscience \cite{morris1981voltage,hodgkin1952quantitative}.

{The bell oscillation is subject to multiple external stimuli, foremost the movement of the surrounding water currents that exert stresses on the bell tissue. Generally, these perturbations impact the oscillatory states of the bell muscles and with these, of the neuronal network, causing a potential shortening or lengthening of the bell oscillation period. This variability in periodicity is captured by the phase coupling function $H_j(.)$ in Eq.~\eqref{eq swarming jelly} \cite{rosenblum2019nonlinear,pikovsky2001synchronization}. Due to the lack of observations and for the sake of simplicity we set $H_j(.) \equiv 0$, assuming no direct response of the bell to the presence of other agents. Future experiments have to be designed to check for such phase couplings and to extract any potential phase coupling from data \cite{rosenblum2001phase}. In Eq.~\eqref{eq swarming jelly}, we indicate the intricate nature of coupling by dependencies on $\mathbf{X}$ and $\boldsymbol \theta$ as the strength of coupling will depend on the position of individuals, the pairwise distance and their orientations.}

\subsubsection*{Orientation dynamics $\theta$}

Jellyfish tend to orient themselves according to different stimuli but they are also subject to the eddy motion induced by their own swimming. The resulting turbulent eddies cause a certain angular drift $L$ which acts to reorient the jellyfish, so that
{\begin{equation}
\dot{\theta}_j = {\cal G}_j(\mathbf{X},\boldsymbol{\theta},{\cal A}_j,F_j,|\mathcal{C}|_j,t) = L_j(t) + G_j(\mathbf{X},\boldsymbol{\theta},{\cal A}_j,F_j,|\mathcal{C}|_j) \; . \label{eq: angular dynamics general}
\end{equation}
$G(.)$ is the angular coupling function.}
The angular diffusion according to $L_j$ is a frequently observed phenomenon on the microscopic \cite{fier2018langevin, saragosti2012modeling} and even mesoscopic \cite{polin2009chlamydomonas} scale. Although the properties of this diffusion have not been investigated yet for jellyfish, we expect a stochastic approximation of local eddies to be valid. This is because the reorientation is realized by eddies that remain, at least temporarily, attached to a jellyfish \cite{gemmell2015control, hoover2017quantifying} before they separate and may lead to communication among different individuals. 

Here, we model the phenomenon by a colored noise  \cite{uhlenbeck1930theory} which allows us to incorporate a simplistic version of correlations and inertial effects of rotation:
\begin{equation}
\dot{L}_j = -\lambda_{\theta} L_j + \eta_j(t), \qquad \langle \eta_j \rangle = 0, \qquad \langle \eta_j\eta'_j \rangle = 2 D({\cal A}_j) \delta(t-t') \; . \label{eq: torque}
\end{equation}
{The parameters $\lambda_{\theta}$ is a measure for the correlation time of the noise and $D$ is the strength of the white noise $\eta_j(t)$. Both, $\lambda_{\theta}$ and $D$ determine the strength of the angular diffusion.} We assume that the resulting motion becomes more prominent when jellyfish search for prey, as individuals start to shuffle water towards their oral arms for feeding purposes \cite{hays2012high, hansson1995behavioural, costello1995flow}. Thus, we allow for an increase of the noise intensity similar to the natural frequencies of agents: 
\begin{equation}
 D({\cal A}_j) = D_{0}[1+\mathcal{R}(f_{\theta},f_{\theta},{\cal A}_j)]\, .
\end{equation}
For simplicity we assume equal response parameters $f_{\theta}=f_{\varphi}$ and we fix the essentially free parameter $D_0 = 0.1 \, \text{rad}^2\ \text{s}^{-3}$, which {determines the angular diffusion in an inactive state (${\cal A}=0$).} 

{The angular coupling function $G(.)$ acts like a potential in the noisy dynamics and causes certain directions to be more favorable than others. In this study, we consider three paradigmatic orientational inputs \cite{albert2011s}: } First, the agents tend to orient themselves such that they counteract the flow in which they are immersed. {Whereas the physiological causes of this behavior are still hypothesized, it is a matter of fact that swimming against the currents increases the survival rate \cite{fossette2015current}. For example, currents may transport jellyfish to undesired regions such as shore lines and swimming against the flow also increases the chances of catching prey}. Second, jellyfish try to avoid regions of turbulent flow in order both to optimize their swimming performances and to escape shear flows that may harm their fragile body. Third, jellyfish orient towards the prey concentration, in particular when under starvation \cite{arai1991attraction}.   
In order to model the interplay of these behaviors, we introduce three separate angular coupling functions $g_j(\theta_j-\delta_{({\bf U}, {\cal C}, F)})$ with respect to the three local directions $\delta_{({\bf U},{\cal C},F)}$, of current, absolute vorticity gradient and prey:
 \begin{equation}
  \tan \left(\delta_{\bf U} \right) = \frac{U_y}{U_x}, \qquad \tan \left(\delta_{|{\cal C}|} \right) = \frac{\partial_y|{\cal C}|}{\partial_x|{\cal C}|},  \qquad \tan \left(\delta_F \right) = \frac{\partial_y F}{\partial_x F}\; .
 \end{equation}
 We assume that the respective angular coupling functions possess a single dominant maximum that can be well approximated by a sinusoidal harmonics \cite{kuramoto2003chemical,rodrigues2016kuramoto, pikovsky2015dynamics, rosenblum2019nonlinear}, resulting in the overall angular dynamics:
\begin{equation}
\dot{\theta}_j = L_j + \varepsilon_{\bf U} \sin \left(\theta_j-\delta_{\bf U} \right) + \varepsilon_{{\cal C}} \sin \left( \theta_j-\delta_{{\cal C}} \right) - \varepsilon_F \sin \left( \theta_j-\delta_{F} \right) \; . \label{eq: orientation state}
\end{equation}
Constructed in this way (for $\varepsilon_{({\bf U},{\cal C},F)} > 0)$, the interplay of external orientation inputs tries to establish a stable orientation in parallel of the tumbling. 
\begin{figure}[!h!]
\centering
\includegraphics[width=0.99\columnwidth, angle=0]{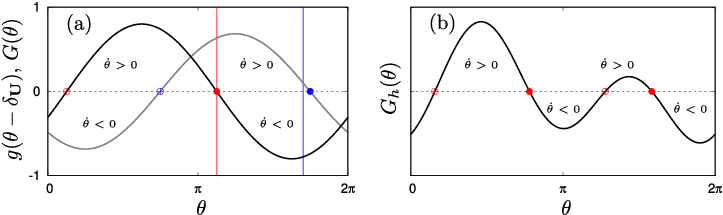}
\caption{{Depicted are examples of the angular dynamics Eq.~\eqref{eq: angular dynamics general} in the absence of noise ($L_j \equiv 0$). (a): Shown in black is the coupling function Eq.~\eqref{eq example coupling U} with $\varepsilon_{\bf U}=0.8$ and $\delta_{\bf U}=\pi/8$. Shown in grey is a generic coupling function Eq.~\eqref{eq: orientation state} with $\varepsilon_{\bf U}=0.1$, $\varepsilon_{\cal C}=0.2$, $\varepsilon_F=0.7$, $\delta_{\bf U}=\pi/8$, $\delta_{\cal C}=6\pi/5$ and $\delta_F=1.7\pi$. Vertical lines indicate the phase points $\theta=\delta_{\bf U} + \pi$ (red) and $\theta=\delta_F$ (blue). Solid dots indicate the stable fixed points and open dots indicate the unstable fixed points of the respective dynamics. (b): Shown is an angular coupling function $G_h(.)$ where $g(\theta- \delta_{\cal C})= \varepsilon_{\cal C} \sin(2(\theta - \delta_{\cal C}))$. The parameters are $\varepsilon_{\bf U}=0.3$, $\varepsilon_{\cal C}=0.5$, $\varepsilon_F=0.1$, $\delta_{\bf U}=\pi/8$, $\delta_{\cal C}=6\pi/5$ and $\delta_F=0.7\pi$. Due to the second harmonics, the dynamics allows for two stable orientations.}}
\label{fig explain phase dynamics}
\end{figure}

{We can check this for example for the simplified angular dynamics 
\begin{equation}
 \dot{\theta}_j= g(\theta - \delta_{\bf U}) = \varepsilon_{\mathbf{U}} \sin(\theta_j-\delta_{\mathbf{U}})\; , \label{eq example coupling U}   
\end{equation}
in which we assume that tumbling is absent ($L_j \equiv 0$) and that jellyfish orient just due to the flow of the water. In this case, there exist exactly two fixed points $\theta^{\star}_j=\delta_{{\bf U}}$ and $\theta^{\star}_j=\delta_{{\bf U}} + \pi$ at which the orientation remains unchanged 
($\dot{\theta}_j=0$, see panel (a) Fig.~\ref{fig explain phase dynamics}). The first point corresponds to swimming in the direction of flow and is by construction unstable. This means that small external perturbation of orientation will grow in this point such that the jellyfish rotate out of alignment. The second point however, is stable and corresponds to swimming against the local direction of flow. For the remaining two orientational inputs, a similar rational applies yielding } 
 $\theta^{\star}_j-\delta_{{\cal C}} = \pi$ (vorticity) or $\theta^{\star}_j-\delta_F=0$ (prey), corresponding to avoidance of turbulence and a directed swimming towards prey respectively. 
 
{The coupling function Eq.~\eqref{eq: orientation state} gives rise to exactly one stable orientation $\theta_j^{\star}$ and its value depends on the coupling parameters $\varepsilon_{({\bf U}, {\cal C}, F)}$ and orientations $\delta_{({\bf U}, {\cal C}, F)}$. For example, when the coupling function $G(.)$ is dominated by coupling to prey ($\varepsilon_{({\bf U}, {\cal C})} \ll \varepsilon_F$), agents mainly orient towards prey such that $\theta_j^{\star} \approx \delta_F$ (grey line in (a) Fig.~\ref{fig explain phase dynamics}). However, it is conceivable that high-order coupling terms are present in the angular dynamics as well. These terms give rise to several permissible orientations of swimming. For example, we can replace $g(\theta - \delta_{\cal C}) = \sin(2(\theta - \delta_{\cal C}))$ in Eq.~\eqref{eq: orientation state} to obtain a high-order coupling function $G_h(.)$ (see panel (b) Fig.~\ref{fig explain phase dynamics}). When noise is present ($L \neq 0$), jellyfish will be perturbed out out their stable orientations. In that case, it depends on the amplitude of the coupling constants how likely it is to stay in the direction of a stable equilibrium point. If the dynamics is multi stable, agents can switch in between different equilibria \cite{StrogatzNLD}. We exploit this fact to introduce directional decision making by allowing for}
parametric switches:
\begin{equation}
\begin{aligned}
 \varepsilon_{\bf U}(F_j) & = \varepsilon_{{\bf U},0}[1-\mathcal{R}(1,f_{\bf U},F_j)], \qquad
\varepsilon_{|{\cal C}|}({\cal A}_j)  = \varepsilon_{|{\cal C}|,0}[1-\mathcal{R}(1,f_{|{\cal C}|},{\cal A}_j)], \\  \varepsilon_F({\cal A}_j) & = \varepsilon_{F,0}\mathcal{R}(1,f_{F},{\cal A}_j)[1-\mathcal{R}(1,f_{g},{\cal A}_j)] \, . \label{eq: responses eps}
\end{aligned}
\end{equation}
Here, $\varepsilon_{({\bf U}, {\cal C},F;0)}$ are the respective angular coupling parameters in inactive (${\bf U},|{\cal C}|$) or active ($F$) state. 
In an inactive state, agents orient against the flow and avoid turbulent regions. An increase of activity will initiate the following cascade of events:
\begin{itemize}
    \item The pulsation frequency $\omega_j({\cal A}_j)$ increases alongside the amplitude of bell strokes and the coefficient of angular diffusion $D({\cal A}_j)$, causing an increase in the speed oscillations of swimming and a more frequent change of the swimming direction.
    \item Jellyfish lose interest in turbulence avoidance. Instead, they initially favor orientations towards high concentrations of prey.
    \item In response to increase in the instantaneous prey concentration, individuals gradually lose their orientation against the flow. 
    \item When a jellyfish reaches high levels of activity, it again loses its interest in swimming towards prey and establishes a state of free floating.
\end{itemize}

Additional influence on the orientation can arise due to the presence of other agents, resulting in further structure formation according to more interaction terms $g_j(\mathbf{x}_k, \mathbf{x}_j, \theta_k, \theta_j, {\cal A}_j)$ \cite{pikovsky2001synchronization,childs2008stability, degond2022environment, ashraf2016synchronization}. To the best of our knowledge, there are no quantitative estimates for a directional interaction function in jellyfish, hence for now we set $G_j(.) \equiv 0$. We keep in mind however that in {dense swarms, probably mostly during the reproduction periods, the mutual interaction between the agents may play a dominant role.}

\subsubsection*{Positional dynamics}

We consider an over-damped positional dynamics of the form:
\begin{equation}
\begin{aligned}
 \dot{\mathbf{x}}_j &= \mathbf{v}_j(\mathbf{X},\boldsymbol{\theta}, \boldsymbol{\varphi}, {\cal A}_j, F_j,\mathbf{U}_j,t) = \mathbf{U}_j + \mathbf{V}_j(\varphi_j,\mathbf{U}_j,{\cal A}_j,\theta_j) \\
 & + \frac{1}{N_j}\sum_{k\neq j}^{N_j} \mathbf{I}_{\text{attr}}(\mathbf{x}_k, \mathbf{x}_j,\varphi_k,\varphi_j, \theta_k, \theta_j) + \mathbf{I}_{\text{rep}}(\mathbf{x}_k, \mathbf{x}_j, \varphi_k,\varphi_j, \theta_k, \theta_j) + \mathbf{F}_{\text{ext}}(t) \; .
\end{aligned}\label{eq: pos state}
\end{equation}
The first two terms on the right hand side account for advection and active  self-propelled motion, whereas the {summations account for the swarm-internal velocity component due to the attraction and repulsion between jellyfish agents}. The last term addresses external forcing mainly due to the effect of confinements or obstacles.  

\begin{figure}[!h!]
\centering
\includegraphics[width=\columnwidth]{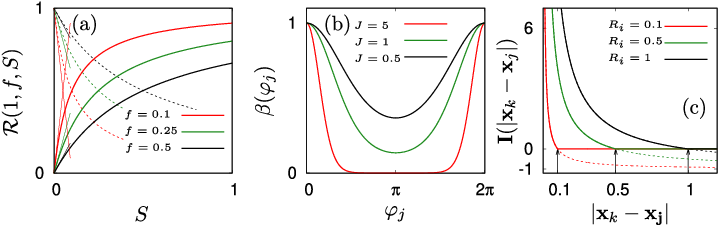}
\caption{{(a): Exemplary response functions Eq.~\eqref{eq: response}. Bold lines correspond to parameter responses of bell frequency $\omega_j({\cal A}_j)$, angular diffusivity $D({\cal A}_j)$ and counter-current speed. Dashed lines indicate the responses of $\varepsilon_{\mathbf{U}}(F_j)$ and $\varepsilon_{\cal C}({\cal A}_j)$ which are given by $1-{\cal R}(.)$. Thin straight lines correspond to the quasi-linear responses with slopes of $f^{-1}$ where parameters are $f \in[0.1,0.25,0.5]$. (b): The bell pulsation function $\beta(\varphi_j)$ for parameter $J \in [5,1,0.5]$. (c): Radial component of the pairwise interaction force Eq.~\eqref{eq attr rep}. Bold lines show the soft-core repulsion used in this study. Dashed lines indicate the omitted attractive part of the force. Black arrows indicate the respective cutoff radii at $R_i=[0.1,0.5,1]$ m.}}
\label{fig: couplings}
\end{figure}

We model the active velocity of a jellyfish by: 
\begin{equation}
\begin{aligned}
    \mathbf{V}_j &= V(\varphi_j,\mathbf{U}_j,{\cal A}_j) \hat{e}(\theta_j) \\
    V(\varphi_j,\mathbf{U}_j,{\cal A}_j) &=  \Big( V_0 + \mathcal{R}(V_a,V_b, |\mathbf{U}_j|+{\cal A}_j) \Big)\beta(\varphi_j) \\
    \beta(\varphi_j) &= \exp\left[J(\cos( \varphi_j) -1)\right] \; . 
\end{aligned}\label{eq: velocity}
\end{equation}
Here, $\hat{e}=[\cos(\theta_j),\sin(\theta_j)]$ is the orientation vector of agent $j$, $V_0$ is the maximal propulsion speed of the jellyfish in their inactive state and parameters $V_a$, $V_b$ characterize speed adaptions due to changes of current and activity. We parameterize the bell oscillation explicitly. Our choice is inspired by the results found in \cite{alben2013efficient, miles2019don, costello2021hydrodynamics, dular2009numerical, park2014simulation, hoover2017quantifying} and many other publications on jellyfish swimming which all report or assume pulsed swimming based on varying types of bell movement. In this paper, we use a simplified symmetric version of bell movement that can be adjusted by {shape} parameter $J$ (see Fig.~\ref{fig: couplings} (b)). When $J$ is small the self propulsion never fully vanishes, as is the case for species that benefit from inertial effects. In contrast, when $J$ is large, jellyfish temporarily come to a halt as can be observed in species that have developed certain types of jet propulsion. 

Attraction and repulsion of agents are taken into account by the terms $\mathbf{I}_{\text{attr}}$ and $\mathbf{I}_{\text{rep}}$, where the dependencies on $\theta$ and $\varphi$ allow to incorporate physiological abilities of single medusae to sense and communicate in a swarm and in a complex environment. Nonetheless, due to the lack of experimental data on these dependencies we keep both $\mathbf{I}_{\text{attr}}$ and $\mathbf{I}_{\text{rep}}$ to be only functions of relative distances. Moreover, since observational evidence for attraction of agents is limited, we assume that the active agents are soft spheres and restrict the interaction to within a radius of $R_i$ (see Fig.~\ref{fig: couplings} (c)) \cite{weeks1971d, aranson2013active, negro2022inertial, degond2022environment, pikovsky2021transition,buttinoni2013dynamical}. Accordingly, $N_j$ is the number of agents found within an interaction distance of $R_i$. The resulting repulsion corresponds to unidirectional sensing and communication:
\begin{equation}
 \mathbf{I}_{\text{attr}}(\mathbf{x}_k, \mathbf{x}_j)= \frac{\mathbf{x}_k-\mathbf{x}_j}{|\mathbf{x}_k-\mathbf{x}_j|}, \qquad \mathbf{I}_{\text{rep}}(\mathbf{x}_k, \mathbf{x}_j) = -R_i \frac{\mathbf{x}_k-\mathbf{x}_j}{|\mathbf{x}_k-\mathbf{x}_j|^2}\; . \label{eq attr rep}
\end{equation}
We denote the resulting summation of contributions in Eq.~\eqref{eq: pos state} by $\mathbf{I}(\mathbf{x}_j)$. 

The type of interaction represents a fluid-like repulsion while studies on active colloids often consider molecular interactions (for instance Lenard-Jones potentials) \cite{rex2007lane, aranson2013active, wysocki2014cooperative, buttinoni2013dynamical}. We choose this approach because jellyfish are soft and most of the time avoid bumping into each other such that the interaction involves only the fluid. As such, the given interaction resembles the leading order terms of a flow field around an active swimmer on the micro scale \cite{giacche2010hydrodynamic, drescher2011fluid}. Thus, {the repulsion} $\mathbf{I}(\mathbf{x}_j)$ only partly represents the turbulent fluid dynamics of jellyfish swimming and we rather see the coupling as a computationally less costly parameterization of the true physical interaction, similar to \cite{o2017oscillators}. 

Finally, the only external force $\mathbf{F}_{\text{ext}}$ accounts for interactions with obstacles. In our case, we mimic the avoidance of walls \cite{albert2011s, hamner1994sun} in a tank. We model the wall forces by a one-dimensional soft-sphere repulsion, similar to Eq.~\eqref{eq attr rep} with a repulsion radius of $0.1$\ m. {A simplistic comparison of the overall model and already existing research on active matter and jellyfish prediction models is shown in Tab.~\ref{table0}.}


\section*{Discussion and results}

We consider three paradigmatic flow environments in a rectangular tank that allow us to compare the performance of our jellyfish model with observational findings. In the first setting, we simulate a cavity flow driven by the movement of the upper wall boundary from the left to the right (see panel (c) of Fig.~\ref{fig: passive agents phase transition}). We fix the wall velocity to $0.4$\ m\ s$^{-1}$ \cite{ghia1982high}. The second setting is a channel flow in a tank of $5$\ m in width and $40$\ m in length (see panel (f) of Fig.~\ref{fig: counter current}). The fluid enters the domain at $+20$\ m and leaves the domain at $-20$\ m. We assume a fully developed Poiseuille flow at the inlet with a magnitude of $-0.045$\ m\ s$^{-1}$ in accordance with experimental parameters in \cite{malul2019levantine}. In the third setting we simulate a cavity flow in a tank of $10$\ m width and $5$\ m length. We let the left and right wall of the tank move upwards at a speed of $0.4$\ m\ s$^{-1}$. This way we generate two counter rotating main gyres that separate the domain dynamically into two parts. Throughout all of our simulations we use fluid parameters given in Tab.~\ref{tab: params inputs}. Thus, the Reynolds numbers of the cavity flows are Re$=10^5$ and for the channel flow it is Re$=4500$.

\begin{table}[!ht]
\begin{adjustwidth}{-0cm}{0cm} 
\centering
\caption{
{\bf Parameters of the jellyfish swarming model Eqs.~(\ref{eq activity state}-\ref{eq: orientation state}) and Eqs.~(\ref{eq: responses eps}-\ref{eq attr rep}) and fluid parameters.}}\label{tab: params inputs}
\begin{tabular}{|l+l|l|l|l|l|l|l|}
\hline
{\bf Phenomenon} & {\bf Parameters [unit]} & {\bf Value} & {\bf External inputs}\\ \thickhline
Attraction and repulsion & $R_i$ [m] & $0.1$ & -- \\
 \hline
 Phase diffusion & $\lambda_{\theta}$ [s$^{-1}$] & $\mathbf{5}$ & -- \\
 & $D_0$ [rad$^2$\ s$^{-3}$] & $0.1$ & \\
 \hline
 Swimming against the flow & $V_0$ [m\ s$^{-1}$] & $\mathbf{0.15}$ & Flow field $\mathbf{U}_j$ \\
  & $V_a$ [m\ s$^{-1}$] & $\mathbf{0.5}$ &  \\
  & $V_b$ [m\ s$^{-1}$] & $\mathbf{0.6}$ &  \\
  & $J$ [--] & $1.0$ &  \\
  & $\varepsilon_{\mathbf{U},0}$ [rad\ s$^{-1}$] & $\mathbf{0.16}$ &  \\
 \hline
 Avoidance of turbulence & $\varepsilon_{{\cal C},0}$ [rad\ s$^{-1}$] & $\mathbf{0.08}$ & Gradient of absolute \\
 & & & Vorticity $\nabla |{\cal C}|_j$ \\
 \hline
 Bell oscillations & $\omega_{j,0}$ [rad\ s$^{-1}$] & $1.2$ & -- \\
 \hline
 Prey search behavior & $\lambda_{\cal A}$ [s$^{-1}$] & $0.005$ & Prey concentration $F_j$ \\
 & $\varepsilon_{F,0}$ [rad\ s$^{-1}$] & $\mathbf{0.16}$ & Prey gradient $\nabla F_j$ \\
 & $f_{\varphi}$ [--] & $0.75$ &  \\
 & $f_{\theta}$ [--], & $0.75$ & \\
 & $f_{\mathbf{U}}$ [$F_0$] & $\mathbf{0.05}$ & \\
 & $f_{\cal C}$ [--] & $\mathbf{0.2}$ & \\
 & $f_{F}$ [--] & $\mathbf{0.2}$ & \\
 & $f_{g}$ [--] & $100$ & \\
\hline
\end{tabular}
\begin{tabular}{|lll|}
\hline
Fluid dynamics parameters \hspace{-0.0cm} & &  \\
\hline
Density $1000$\ kg\ m$^3$ & Effective viscosity $0.4$\ Ns\ m$^{-2}$ &  \\
Prey diffusivity $D_F=0.001$\ m$^2$\ s$^{-1}$ & & \hspace{2.07cm} \\
\hline
\end{tabular}
\vspace{0.1cm}
\begin{flushleft} 
Parameters estimated in this study are shown in bold letters. The fluid time step is $0.001$\ s. Agent and fluid variables are stored every $0.1$\ s.
\end{flushleft}
\label{table1}
\end{adjustwidth}
\end{table}

\subsection*{Wall effects}

{Here, we first discuss what effects arise due to repulsive interactions at walls (${\bf F}_{\text{ext}}$) and due to agent-agent interaction (${\bf I}(\mathbf{x}_j)$ in Eq.~\eqref{eq: pos state}. For this, we simulate two ensembles with reduced dynamics }
\begin{equation}
 \dot{\mathbf{x}}_j = \mathbf{U}_j + \mathbf{F}_{\text{ext}}, \qquad \text{and} \qquad
 \dot{\mathbf{x}}_j = \mathbf{U}_j + \mathbf{I}(\mathbf{x}_j) + \mathbf{F}_{\text{ext}} \; . \label{eq pass rep tracers}
\end{equation}
{The first ensemble mimics passive tracers. They get advected by their local fluid velocities $\mathbf{U}_j$ and they are confined by the wall forces ${\bf F}_{\text{ext}}$. In the second ensemble, also volume exclusion is present due to $\mathbf{I}(\mathbf{x}_j)$. It is well known that in the presence of some volume-excluding agent-agent interactions, states of (quasi-) periodic spacial order may emerge \cite{negro2022inertial,strandburg1988two, gasser2010melting}. We seek to avoid such states as they would mask the structure formation process due to swimming and orientation in our simulations. In our model, the transition towards ordered states is influenced by the interaction radius $R_i$ and by the number of agent in the domain, which we have to select accordingly. 
}

\begin{figure}[!h!]
\centering
\includegraphics[width=\columnwidth]{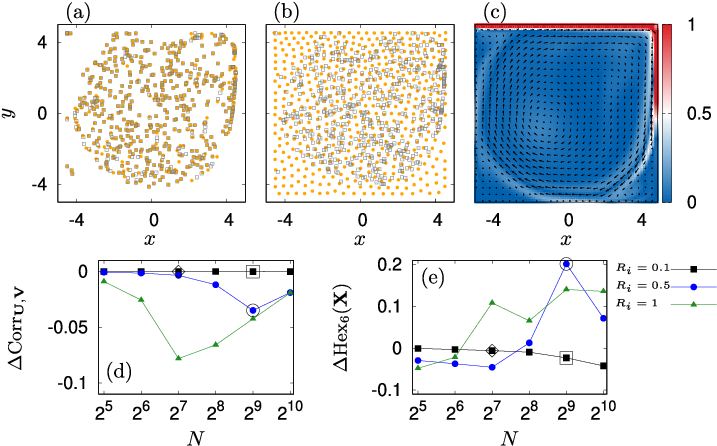}
\caption{Results for a cavity flow at Re$=10^5$ and for $N=512$ agents. The upper wall moves at a constant speed of $0.4$ m\ s$^{-1}$ to the right. (a): Swarm of repelling jellyfish (dots) for $R_i=0.1$\ m. Passive particles are indicated by squares and almost overlap with the position of their active counterparts (orange). (b): A similar swarm but now for $R_i=0.5$\ m. In this setting the swarm is heavily influenced by the confinement. (c): Flow field $\mathbf{U}(\mathbf{x},t)$ (arrows) and absolute vorticity $|{\cal C}|(\mathbf{x},t)$ (values higher than unity are cut off) on a domain of $10$\ m $\times 10$\ m. (d): Difference in velocity correlation of active and passive swarm for $R_i\in[0.1,0.5,1]$\ m (squares, dots, triangles). (e): The corresponding difference of hexatic order parameters {for averaging over the $6$ nearest neighbors}. Marked squares and circles correspond to panel (a) and (b) respectively. Diamonds mark the parameter setting chosen in this paper.}
\label{fig: passive agents phase transition}
\end{figure}

First, we set the interaction distance at walls to $0.5$\ m. Example results are depicted in Fig.~\ref{fig: passive agents phase transition}. We calculate the correlations of velocity and the hexagonal order parameter 
\begin{equation}
    \text{Corr}_{\mathbf{V},\mathbf{U}} =  \frac{1}{2} \Big[ \text{Corr}_{\dot{x},\mathbf{U}_x} +     \text{Corr}_{\dot{y},\mathbf{U}_y} \Big], \qquad     \text{Hex}_n(\mathbf{X}) = \overline{\; \Big| \overline{\exp(i6\delta)}_{n} \Big|\; }_N \label{eq ex order def}
\end{equation}
for the active and passive tracers.  {To obtain the local hexagonal order, we find for each agent $j$ its $n$ nearest neighbors. Then, $\delta$ is the relative angle between agents $j$ and $k$ (obeying $\tan(\delta_{j.k}) = (y_j-y_k)/(x_j-x_k)$) and the complex order parameter is given by an average of $\exp(i6\delta_{j,k})$ over these $n$ nearest neighbors. To obtain the ensemble order, we finally} average over the ensemble of $N$ agents \cite{schmidle2012phase, gasser2010melting}. The velocity correlation measures how much the ensemble follows the background flow while the hexagonal order parameter quantifies local lateral order. {For example if the agents were to assemble in perfect hexagons, angles of the $6$ nearest neighbors $\delta_{j,k}=\pi/6$ inside the domain. Accordingly, $\text{Hex}_6(\mathbf{X})=1$. If any other symmetry would arise or any other disordered state, it would be indicated by a certain lower level of $\text{Hex}_6(\mathbf{X})$.} 

We see that the swarm undergoes a phase transition from almost free floating to a disordered crystal phase (panels (a) and (b) in Fig.~\ref{fig: passive agents phase transition}), depending on the parameters $R_i$ and $N$. This can be observed in Fig.~\ref{fig: passive agents phase transition} (d,e) for the time averages of $\Delta \text{Corr}_{\mathbf{U},\mathbf{V}} = \langle \langle \text{Corr}_{\mathbf{U},\mathbf{V}} - \text{Corr}_{\mathbf{U},\mathbf{V};\text{passive}}\rangle \rangle_t$ and $\Delta \text{Hex}_6(\mathbf{X}) = \langle \langle \text{Hex}_6(\mathbf{X}) - \text{Hex}_{\text{passive},6}(\mathbf{X})\rangle \rangle_t$ after an initial transient of $50$\ s. 

At small swarm sizes the agents essentially follow inertial trajectories. When the number of agents increases, we see that for larger interaction radii the velocity correlation drops due to swarm-internal repulsive movements. Depending on $R_i$, the particles essentially lose their ability to float freely. However, at sufficiently large area densities, agents are so closely packed that local repulsive interactions cancel each other. Since the positional dynamics Eq.~\eqref{eq pass rep tracers} is over-damped, it means that particles are still tied to their local fluid velocities such that their motion gains back its correlation with the local flow \cite{strandburg1988two}. 

In parallel, hexagonal order reveals that for $R_i=0.1$\ m, passive and repulsive swarm behave essentially similarly. At high area densities, we observe that repulsive movement causes a decrease in relative order. On the contrary, we see that the order shows a maximum before it decreases again for larger repulsion radii. This is because the natural increase in hexatic symmetry is disrupted by the flow perturbations and by the quadratic symmetry of the boundary which infiltrates into the domain once long range correlations establish due to dense packing.

Based on this discussion, we set {the number of agents} $N=128$ and {the repulsion radius} $R_i=0.1$\ m as these parameters guarantee almost free floating of the jellyfish-like particles.

\subsection*{Swimming against the flow}

We are in particular interested in the swimming dynamics of the jellyfish \textit{Rhopilema nomadica}. For this species, the ability to counteract the underlying current has been measured in a long channel at different inflow velocities \cite{malul2019levantine}. The swimming speed in the absence of a flow was estimated to $0.067$\ m\ s$^{-1}$ and an active swimming speed of $0.083$\ m\ s$^{-1}$ was reported for an inlet velocity of $0.045$\ m\ s$^{-1}$. 

\begin{figure}
\centering
\includegraphics[width=\columnwidth]{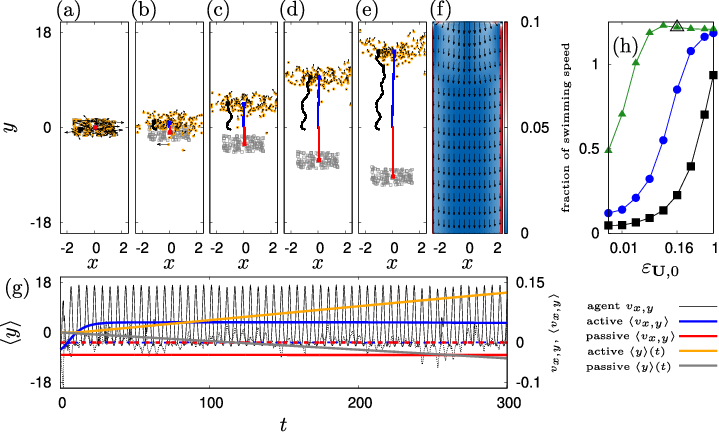}
\caption{Simulation results of counter-current swimming in a horizontal channel flow at Re$=4500$. The peak inflow velocity is set to $0.045$\ m\ s$^{-1}$. Positions of active (orange) and passive (grey) agents at (a): $t=0$\ s, (b): $t=30$\ s, (c) $t=100$\ s, (d): $t=200$\ s and (e): $t=300$\ s for $\varepsilon_{\mathbf{U},0}=0.16$ rad\ s$^{-1}$, $\lambda_{\theta}=5$\ s$^{-1}$, $D_0=0.1$\ rad$^2$\ s$^{-3}$, $J=1$, $V_0=0.15$\ m\ s$^{-1}$, $V_a=0.5$\ m\ s$^{-1}$ and $V_b=0.6$\ m\ s$^{-1}$. Velocities of active agents are indicated by arrows. An exemplary agent trajectory is shown in black. Blue and red trajectories indicate the center of mass movement of active and passive agents. (f): Flow field $\mathbf{U}(\mathbf{x},t)$ (arrows) and absolute vorticity $|{\cal C}|(\mathbf{x},t)$ (color, values are cut off at $0.1$\ rad\ s$^{-1}$) on a lattice of $41 \times 321$ points ($5$\ m $\times 40$\ m) (g): Averaged trajectories $\langle y \rangle(t)$ (left ordinate) for passive (grey) and active (orange) agents. Corresponding velocities are shown on the right ordinate. Dashed lines indicate the $x$-components of velocity, bold lines show the $y$-components. {The thin black lines indicate the velocities of the exemplary agent in panels (a-e).} (h): Fraction of swimming speed as a function of the orientation strength for $D_0=0.1$\ rad$^2$\ s$^{-3}$ and $\lambda_{\theta} \in [0.2,1,5]$ s$^{-1}$ in black, blue and green respectively. The optimal parameter set is indicated by a bigger triangle.} \label{fig: counter current}
\end{figure}

We adapt the model of passive-repulsive tracers Eq.~\eqref{eq pass rep tracers} by allowing for an angular dynamics affected by angular diffusion and orientation against the flow in Eq.~\eqref{eq: orientation state}, a non-zero swimming speed according to Eq.~\eqref{eq: velocity} and bell pulsation Eq.~\eqref{eq: internal state}, resulting in the model:
\begin{equation}
    \begin{aligned}
        \dot{\mathbf{x}}_j = \mathbf{U}_j + V(\varphi_j,\mathbf{U}_j)\hat{e}(\theta_j) + \mathbf{I}(\mathbf{x}_j) + \mathbf{F}_{\text{ext}},\\
        \dot{\varphi}_j = \omega_{j,0}, \qquad \dot{\theta}_j = L_j + \varepsilon_{\mathbf{U},0} \sin(\theta_j-\delta_u) \; .
    \end{aligned} \label{eq counter curr}
\end{equation}
Panels (a-e) Fig.~\ref{fig: counter current} show snapshots of the resulting swarm dynamics.

The, swimming dynamics is defined by the parameters {$\lambda_{\theta}$ (inverse of the correlation time for the angular noise), $D_0$ (amplitude of the angular white noise), $V_0$ (amplitude of swimming speed in the inactive state), $V_a$, $V_b$ (velocity response parameters), $\varepsilon_{\mathbf{U},0}$ (angular coupling constant) and $J$ (shape parameter of velocity oscillations)}. Here we adjust parameters of the model to experimental results \cite{malul2019levantine}: We choose $J=1$ to mimic inertia of medusae during a bell stroke (see Fig.~\ref{fig: couplings} (a)). To obtain other parameters, we assume that agents are fully aligned against the flow and we average the orientation dynamics over time. Then, the {oscillatory component $\beta(\varphi)$} results in a factor of roughly $0.47$ and other parameters follow directly: $V_0=0.15$\ m\ s$^{-1}$, $V_a=0.5$\ m\ s$^{-1}$ and $V_b=0.6$\ m\ s$^{-1}$. 

The corresponding averaged swimming velocity is $0.087$\ m\ s$^{-1}$. We compensate the excess by allowing for angular diffusion in the simulations. Then, the dynamics depends on parameters $\varepsilon_{\mathbf{U},0}$, $\lambda_{\theta}$ and $D_0$. We fix $D_0=0.1\ \text{rad}^2\ \text{s}^{-3}$ and let $\lambda_{\theta} \in [0.2,1,5]$\ s$^{-1}$ {($\sigma^{L} = \sqrt{D_0/\lambda_{\theta}} \in [0.71,0.32,0.14]$\ rad\ s$^{-1}$)}. We perform a parameter scan for $\varepsilon_{\mathbf{U},0} \in[0.005\cdot2^{(0,1,\ldots 7)},1]$\ rad\ s$^{-1}$ and compute the time average of $|\langle \dot{\mathbf{x}} \rangle-\mathbf{U}|$ in the central two third of the simulation time (see panel (g) Fig.~\ref{fig: counter current}). We divide the average by $0.067$\ m\ s$^{-1}$ to obtain the fraction of swimming speed in presence of a current, similar to the analysis in \cite{malul2019levantine}. Panel (h) in Fig.~\ref{fig: counter current} depicts the transition from noisy advection to active counter-current swimming. Our agents swim $21\%$ faster against the flow at $\varepsilon_{\mathbf{U},0}=0.16$\ rad\ s$^{-1}$ and for $\lambda_{\theta} = 5$\ s$^{-1}$ (Fig.~\ref{fig: counter current} bigger triangle in panel h). 

The essence of the counter-current swimming can be understood by means of the angular dynamics in Eq.~\eqref{eq counter curr}. We see that the noise has a standard deviation of $\sigma_L$ such that we expect the existence of the fixed point solution $\theta_j^{\star}-\delta_u = \pi$ to become more likely when $\varepsilon_{\mathbf{U},0}>\sigma^L$. For the given parameter, this critical average coupling is $\sigma^L=0.14$ rad\ s$^{-1}$ which indeed is slightly beyond the transition zone of panel (h) Fig~\ref{fig: counter current}. In turn, phase slips across the fixed point solution when the noise is temporarily too strong, can still occur. Then, the agents start to tumble or even rotate for some time such that their effective propulsion speed against the flow is drastically reduced. From panel (h) Fig.~\ref{fig: counter current} it can be seen that such events are unlikely before the fixed point threshold is reached. Thus, we consider the reconstructed set of parameters to be optimal as it stipulates the frequently encountered phenomenon of \textit{criticality} in neuronal responses and biological systems in general \cite{hesse2014self,mora2011biological}.

\subsection*{Turbulence avoidance and structure} \label{sec vortex avoidance}

\begin{figure}[!h!]
\centering
\includegraphics[width=\columnwidth]{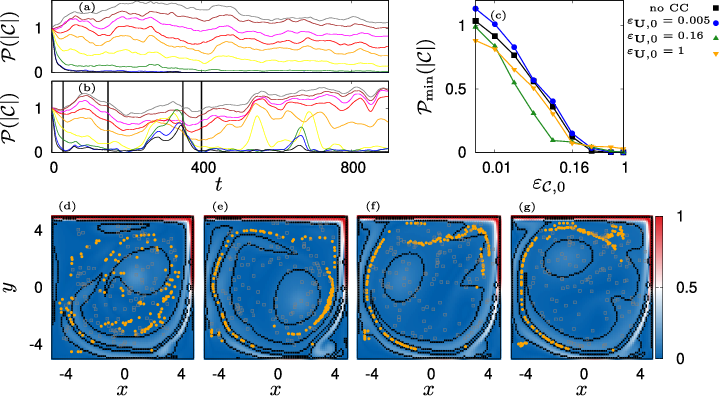}
\caption{Depicted are the time courses of turbulence fraction ${\cal P}(|{\cal C}|)$ given (a): $\varepsilon_{\mathbf{U},0}=0.005$\ rad\ s$^{-1}$ and (b): $\varepsilon_{\mathbf{U},0}=1$\ rad\ s$^{-1}$ for $\varepsilon_{{\cal C},0} \in [0.005\cdot 2^{(0,\ldots, 7)},1]$\ rad\ s$^{-1}$ in grey, brown, magenta, red, orange, yellow, green, blue and black respectively. Panel (c) depicts the averaged baseline turbulence fraction for four modes of counter-current swimming. The flow domain and the position of active agents for $\varepsilon_{{\cal C},0}=0.32$ is shown in (d): $t=30$\ s, (e): $t=150$\ s, (f): $t=350$\ s, (g): $t=400$\ s, also indicated by black vertical lines in panel (b). Characteristic regions are indicated by black boundaries. The vorticity is indicated with a color scale.} \label{fig: vortex avoidance}
\end{figure}

The avoidance of turbulent regions in a flow can be implemented in a straight forward way by a second alignment term in Eq.~\eqref{eq counter curr} of strength $\varepsilon_{{\cal C},0}$:
\begin{equation}
    \dot{\theta}_j = L_j + \varepsilon_{u,0} \sin(\theta_j-\delta_u) + \varepsilon_{{\cal C},0} \sin(\theta_j-\delta_{\cal C}) \; .
\end{equation}
This second term causes the agents to also orient against the gradient of absolute vorticity and thus to swim away from regions of turbulence. As a consequence, flow structures are imprinted on the swarm orientation and can cause the formation of filaments and patches (see panels (d-g) in Fig.~\ref{fig: vortex avoidance}).

We quantify the success in avoiding turbulent regions by means of the indication number Eq.~\eqref{eq: pearson} and define the fraction: 
\begin{equation}
    {\cal P}(|{\cal C}|) = \frac{{\cal N}(|{\cal C}|)}{ {\cal N}(|{\cal C}|_{\text{passive}})} \; . 
\end{equation}
When the agents all avoid turbulence, ${\cal P}(|{\cal C}|) = 0$, when agents are initially randomly distributed and the active agents all reside inside areas of high turbulence, ${\cal P}=1/{\cal N}(|\cal C|_{\text{passive}})$. The characteristic regions where $|{\cal C}|>\overline{|{\cal C}|}_{\text{passive},N}$ are indicated by the black boundary in Fig.~\ref{fig: vortex avoidance}. We see that these regions for the cavity flow evolve essentially into two substructures, an elliptic central region and a long and narrow hose-like structure generated by the clockwise main current.

Mainly influenced by these flow structures, there appear essentially two classes of transients. First, when counter-current swimming is switched off or if the orientation against the flow is weak ($\varepsilon_{\mathbf{U}}=0.005$\ rad\ s$^{-1}$), the number of agents in turbulent regions decreases with increased $\varepsilon_{\cal C}$ and the transition out of turbulence takes place essentially in a short initial time interval (see Fig.~\ref{fig: vortex avoidance} panels (a)). Second, when counter-current swimming is sufficiently strong, the jellyfish effectively stand still in the water or achieve a net velocity against the flow. In that situation, they stay in a surrounding turbulence structure for a prolonged period of time and can group along this structure into patches or long filaments (see Fig.~\ref{fig: vortex avoidance} panels (d-g)). Nevertheless, the swarm can eventually encounter turbulent regions in which it stays for some time before it starts avoiding these regions (see Fig.~\ref{fig: vortex avoidance} panel (f)). When this happens, we observe intermittent spikes in ${\cal P}(|{\cal C}|)$ that decrease in amplitude before the swarm has adjusted to the turbulence structures (see Fig.~\ref{fig: vortex avoidance} panel (b)).

Since spikes are isolated and transient events in time, we first omit the initial $50$\ s and calculate the maximal and minimal fractions $\hat{\cal P}(|{\cal C}|)$ and $\tilde{\cal P}(|{\cal C}|)$. Then, we take into account only values ${\cal P}(|{\cal C}|)<0.9\tilde{\cal P}(|{\cal C}|)+0.1\hat{\cal P}(|{\cal C}|)$ to obtain the time average ${\cal P}_{\text{min}}(|{\cal C}|)$ (see Fig~\ref{fig: vortex avoidance} panel (c)). The resulting curves show that the baseline avoidance of turbulence is essentially independent of $\varepsilon_{\mathbf{U},0}$.

Additionally, we characterize the pattern formation by {$\Delta \text{Hex}_{12}(\mathbf{X}) = \langle \langle \text{Hex}_{12}(\mathbf{X}) - \text{Hex}_{\text{passive},12}(\mathbf{X})\rangle \rangle_t$} and the difference of standard deviation $\Delta \Sigma = \langle \langle \Sigma(\mathbf{X}) - \Sigma_{\text{passive}}(\mathbf{X}) \rangle \rangle_t$ where 
\begin{equation}
 \Sigma(\mathbf{X}) =  \sqrt{\frac{(\sigma^x)^2}{2 (\sigma^x)^2(0)} + \frac{(\sigma^y)^2}{2 (\sigma^y)^2(0)}} \; . \label{eq: sigma}
\end{equation}
Here, $\sigma^x$ and $\sigma^y$ are the standard deviations of the agents positions in $x$ and $y$-direction respectively at time $t$. At $t=0$, $\Sigma(\mathbf{X})=1$ and stays approximately constant for the passive swarm while it increases or decreases according to the evolution of the active swarm. {We simulate the dynamics for $900$\ s and take into account values in the last $300$\ s for time averaging. In contrast to the analysis of regular structural order, we have chosen $12$ nearest neighbors for averaging of the phase factor $\exp(i6\delta)$ in Eq.~\eqref{eq ex order def}}, because it allows us to better resolve the effect of filamentation.

\begin{figure}[!ht!]
    \centering
    \includegraphics[width=\columnwidth]{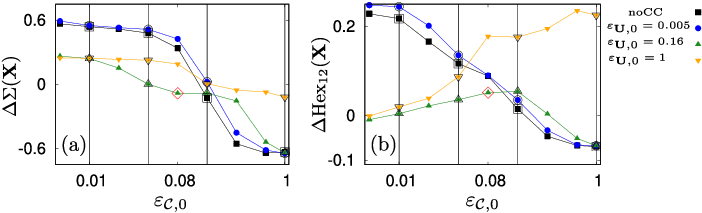}
    \caption{Measures of structure formation. (a): Spread of the swarm $\Delta \Sigma(\mathbf{X})$, (b): Local hexagonal order $\Delta \text{Hex}_{12}(\mathbf{X})$ for four settings of counter-current swimming. Shown as vertical lines are columns in Fig.~\ref{fig: vortex avoidance scatter plots} for $\varepsilon_{{\cal C},0}\in[0.01,0.04,0.16,1]$\ rad\ s$^{-1}$. The red diamond indicates the optimal coupling constant for turbulence avoidance ($\varepsilon_{\bf U}=0.16$\ rad\ s$^{-1}$, $\varepsilon_{\cal C}=0.08$\ rad\ s$^{-1}$).}
    \label{fig: jelly structure formation}
\end{figure}

\begin{figure}[!h!]
\centering
\includegraphics[width=\columnwidth]{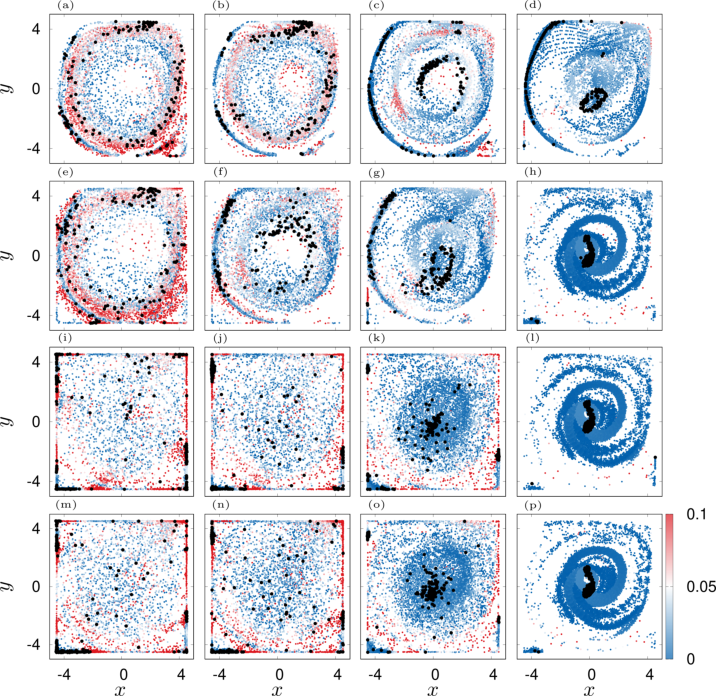}
\caption{Scatter plots of agents position over $900$\ s. Color indicates $|\cal C|$. The final configurations are shown with black dots. Panels (a-d): $\varepsilon_{\mathbf{U},0}=1$, (e-h): $\varepsilon_{\mathbf{U},0}=0.16$, (i-l): $\varepsilon_{\mathbf{U},0}=0.005$, (m-p): no counter-current swimming. Columns from left to right correspond to $\varepsilon_{{\cal C},0}=[0.01,0.04,0.16,1]$ respectively.} \label{fig: vortex avoidance scatter plots}
\end{figure}

We observe that the structure formation process heavily depends on the combination of $\varepsilon_{{\cal C},0}$ and $\varepsilon_{\mathbf{U},0}$. When swimming against the flow is switched off or if it is weak, the jellyfish follow scattered trajectories and mostly get trapped inside the corners of the domain, causing a large spread and a relatively high order of the swarm due to compression at the walls (Fig.~\ref{fig: jelly structure formation} black and blue curves and Fig.~\ref{fig: vortex avoidance scatter plots} panels (i, j, m, n)). 


\begin{figure}[!h!]
\centering
\includegraphics[width=\columnwidth]{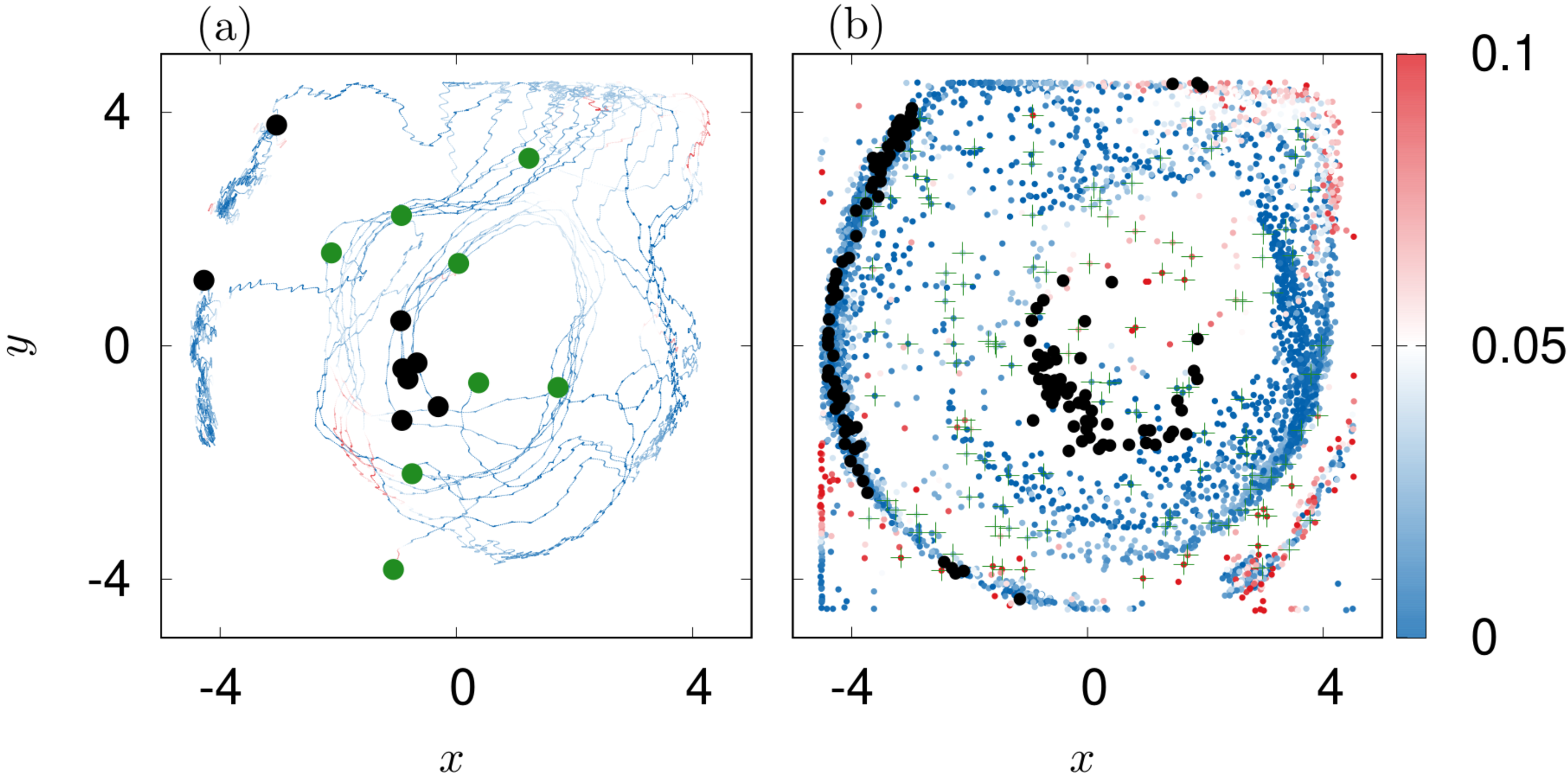}
\caption{Optimal swarming dynamics for $\varepsilon_{\bf U}=0.16$\ rad\ s$^{-1}$, $\varepsilon_{\cal C}=0.08$\ rad\ s$^{-1}$. (a): starting positions (green) and final positions (black) of 8 exemplary jellyfish. The trajectories of the particles are shown in color. The color code indicates the encountered amplitude of absolute vorticity. (b): Scatter plot of all particles similar to Fig.~\ref{fig: vortex avoidance scatter plots}. Additionally shown are green crosses for the position of the agents at $t=0$\ s.} \label{fig: vortex avoidance scatter plots optimal case}
\end{figure}
With increased $\varepsilon_{{\cal C},0}$ the swarm transitions into a single patch in the center (Fig.~\ref{fig: vortex avoidance scatter plots} panels (h, k, l, o, p)) which has a minimal spread and a minimal order. The latter is a consequence of the constant regrouping due to agents that bump into the cluster and advection. On the contrary, when the agents orient more against the flow, they are able to group into long lasting and coexisting coherent structures, namely filaments (Fig.~\ref{fig: vortex avoidance scatter plots} (a-g)) and rings (Fig.~\ref{fig: vortex avoidance scatter plots} (a,b,e,f)). Most prominent is the transition for the optimal, $\varepsilon_{\mathbf{U},0}=0.16$\ rad\ s$^{-1}$. For this setting, the swarm undergoes a transition from ring patterns over filaments towards patches. This transition is indeed signified by two jumps in the spread $\Delta \Sigma$ and an increase of local order before the system settles into the disordered patch ({see green lines in Fig.~\ref{fig: jelly structure formation}}). In contrast, for $\varepsilon_{\mathbf{U},0}=1$\ rad\ s$^{-1}$, the swarm transitions later from ring patterns into filaments. Based on this investigation, we chose $\varepsilon_{\cal C}=0.08$\ rad\ s$^{-1}$ {($\varepsilon_{\bf U}=0.16$\ rad\ s$^{-1}$) as an optimal pair of coupling values (red diamond in Fig.~\ref{fig: jelly structure formation}, see also Fig.~\ref{fig: vortex avoidance scatter plots optimal case}). It gives rise to a dynamics that sits ''in between'' scattered states and patches and thus, ensures a relatively rich behavior of the simulation model as already small variations of $\varepsilon_{\cal C}$ will lead to a significant change in the swarming patterns. We again favor such a behavior as it mimics the frequently encountered phenomenon of \textit{criticality} in biological systems \cite{hesse2014self,mora2011biological}}

\subsection*{Foraging}

\begin{figure}[!h!]
    \centering
    \includegraphics[width=\columnwidth]{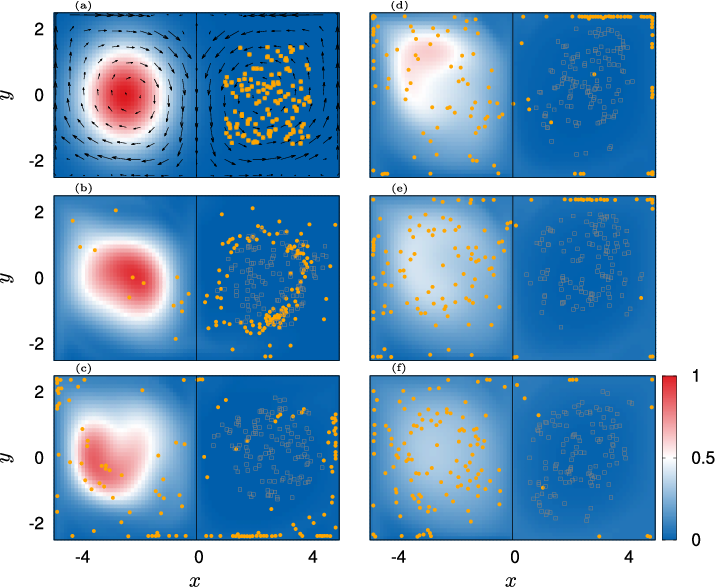}
    \caption{Depicted are snapshots of an optimal search behavior. The model is simulated with parameters found in Tab.~\ref{tab: params inputs}. Arrows in panel (a) show the flow field. Orange dots indicate the actively swimming jellyfish. Squares indicate the passive tracers. The color map indicates the concentration of prey in the water. (a): $t=0$\ s, (b): $t=30$\ s, (c): $t=100$\ s, (d): $t=300$\ s, (e): $t=600$\ s, (f): $t=900$\ s.}
    \label{fig: tracer optimal search}
\end{figure}

Our final goal is to capture the principal behavior of foraging based on the full dynamics Eqs.~(\ref{eq activity state}-\ref{eq: orientation state}) and Eqs.~(\ref{eq: responses eps}-\ref{eq attr rep}). We simulate the swarm in a double-gyre flow in which we initialize the agents in the right gyre at $x>0$ and the prey in the left gyre at $x<0$ (see Fig.~\ref{fig: tracer optimal search} panel (a)). Thus, in this setting the jellyfish are dynamically separated from prey by a strong central current at $x=0$ \cite{arai1991attraction}. We start our model with the previously obtained optimal parameter setting of Tab.~\ref{tab: params inputs} for counter-current swimming and turbulence avoidance. Overall, each model instance runs for $900$\ s. 

\subsubsection*{Response mechanisms}

\begin{figure}[!h!]
    \centering
    \includegraphics[width=\columnwidth]{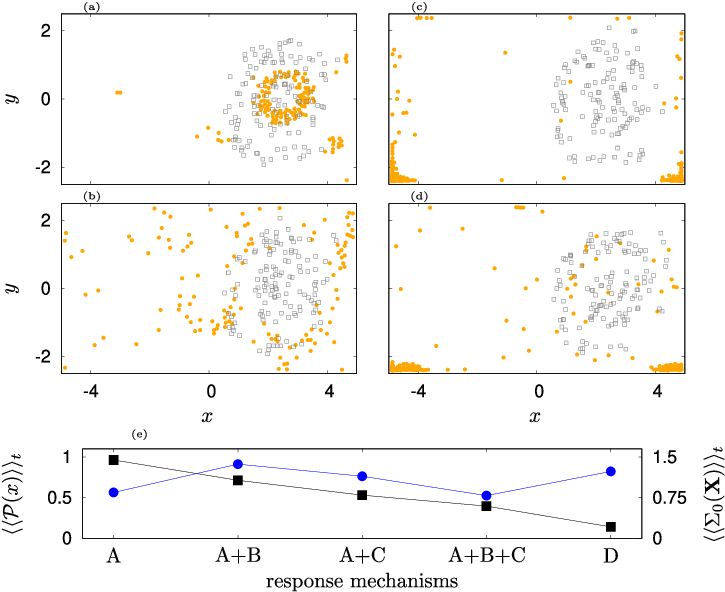}
    \caption{Depicted are the positions of jellyfish (orange dots) and passive tracers (grey squares) due to response mechanisms (a): A, (b): A+B, (c): A+C, (d): A+B+C. {Snapshots depict the swarm in the end of the covered simulation time of $900$\ s. In comparison, the optimal full response is shown for scenario D in Fig.~\ref{fig: tracer optimal search}}. Panel (e) depicts the time average of agent fraction $\langle \langle {\cal P}(x) \rangle \rangle_t $ that remains in the right of the tank (black squares) and the time average of the spreading of agents in the left part of the tank $ \langle \langle \Sigma_0(\mathbf{X}) \rangle \rangle_t$ (blue dots).}
    \label{fig: response mechanisms}
\end{figure}

The turbulence avoidance causes the swarm to form a ring structure in the right gyre of the flow (see Fig.~\ref{fig: tracer optimal search} panel (b) and Fig.~\ref{fig: response mechanisms} panel (a)). We check first the impact of subsequent combinations of response mechanisms on the ability to search for prey:
\begin{itemize}
    \item[A] Higher pulsation frequency $\omega({\cal A})$ and a larger angular diffusion $D({\cal A})$.
    \item[B] Reduction of turbulence avoidance by means of a response $\varepsilon_{\cal C}({\cal A})$. We set {the slope parameter for $\varepsilon_{\cal C}$ to $f_{\cal C}=0.02$, allowing for an almost immediate reduction when activity increases.}
    \item[C] Bi-linear dependence of propulsion velocity $V(\mathbf{U}_j+{\cal A}_j)$.
    \item[D] The full response dynamics at optimal parameters of Tab.~\ref{tab: params inputs}. (This search dynamics is depicted in Fig.~\ref{fig: tracer optimal search}.)
\end{itemize}
Prey search by jellyfish is variable but there exists experimental evidence \cite{arai1991attraction, hays2012high} and observational experience \cite{zafrirpersonal} that can be condensed to the following two performance criteria for our tank simulation: {When jellyfish is searching successfully for prey, }
\begin{itemize}
    \item The fraction of jellyfish in the right {(oligotropic)} gyre is minimal, {meaning that the majority of individuals manages to cross into the left domain at $x<0$ or that individuals group in regions where prey is found.}.
    \item The jellyfish spread evenly in the left gyre once they have crossed the barrier. {This mimics the free floating state when individuals shuffle water to the oral arms.}
\end{itemize}
We quantify the crossing performance by means of the fraction
\begin{equation}
    {\cal P}(x) = \frac{{\cal N}_0(x)}{{\cal N}_0(x_{\text{passive}})}\; ,
\end{equation}
where we indicate by index $0$ that only jellyfish in the right gyre ($x>0$) contribute ({$\hat{x}=0$}). Accordingly, {a fraction} ${\cal P}(x)=1$ {indicates that} all jellyfish and all passive agents are found in the right gyre and it is zero, when all jellyfish have crossed the flow barrier. To quantify the spread of the swarm {after the crossing of the barrier}, we calculate $\Sigma_0(\mathbf{X})$ {according to Eq.~\eqref{eq: sigma}. The index $0$ indicates that for this averaging only jellyfish at $x<0$ are taken into account}.

While frequency and diffusion responses (A) do not result in significant searching (Fig.~\ref{fig: response mechanisms} panel (a)), already enabling weakening of the turbulence avoidance (B) allows some agents to cross the flow barrier and to spread partly in the left gyre (Fig.~\ref{fig: response mechanisms} panel (b)). A velocity response to activity increase (C) leads to a further increase in the number of crossings. However, jellyfish get trapped in the corners of the tank because direction is no longer chosen based upon turbulence avoidance {and higher velocities increase the chance of bumping into a wall where the movement is constrained.}(Fig.~\ref{fig: response mechanisms} panel (c,d,e))). 

\subsubsection*{Interplay of directed search and counter-current swimming}

Next, we also allow for an orientation into the direction of the prey. We find that this additional orientation ($\delta_F$) leads to complex decision making in our jellyfish model based on  an interplay of directed prey search and counter-current swimming. We start our analysis by limiting the set {of prey angular-coupling constant and response parameters} $[\varepsilon_{F,0}, f_{\cal C}, f_{F}, f_{\mathbf{U}},f_{g}]$ to symmetric switching from turbulence avoidance to prey searching ($f_{F}=f_{\cal C}$). And, we fix {the ignorance parameter} $f_{g}=100$, which means that even after prolonged periods of preying, resulting in high activity ${\cal A}$, the jellyfish {remain ''greedy'' for food}. This leaves us with $\varepsilon_{F,0}$, $f_{F}$ and $f_{\mathbf{U}}$. 

\begin{figure}[!h!]
    \centering
    \includegraphics[width=\columnwidth]{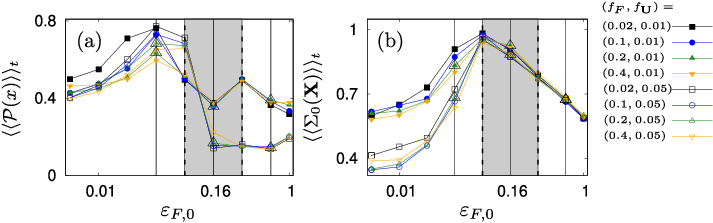}
    \caption{Depicted are the time averaged fraction of agents at $x>0$, $\langle \langle {\cal P} (x) \rangle \rangle_t$ in panel (a) and the spreading $ \langle \langle \Sigma_0(\mathbf{X}) \rangle \rangle_t$ of agents at $x<0$ in panel (b). Empty and filled markers correspond to $f_{\mathbf{U}}=0.05 F_0$ and $f_{\mathbf{U}}=0.01 F_0$ respectively. The grey box indicates the region of favorable $\varepsilon_{F,0}$\ rad\ s$^{-1}$. Vertical black lines and larger triangles indicate parameters used in Fig.~\ref{fig: jelly tracer two clusters}.}
    \label{fig: jelly tracer spread fraction}
\end{figure}

\begin{figure}[!h]
    \centering
    \includegraphics[width=\columnwidth]{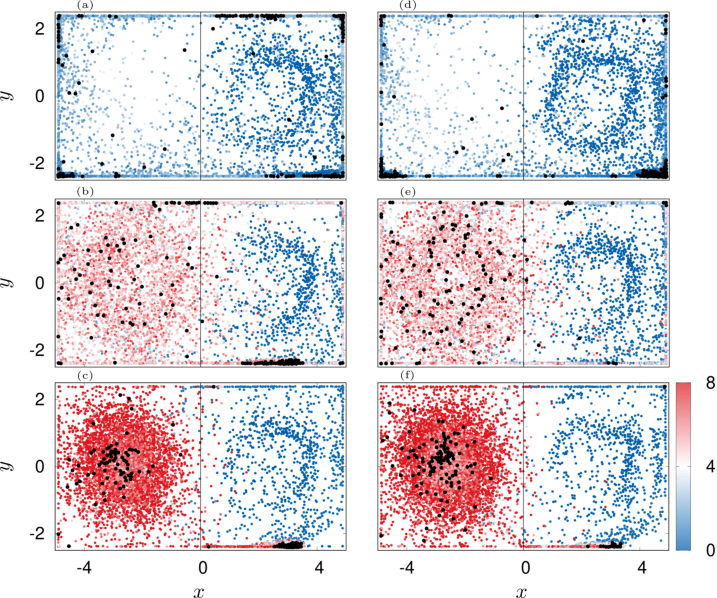}
    \caption{Depicted are scatter plots of jellyfish positions over $900$\ s and with an increment of $13$\ s for parameter $f_F=0.2$. Color indicates the activity ${\cal A}$. Black dots indicate swarm in the end of the simulation. (a): $f_{\mathbf{U}}=0.01 F_0$, $\varepsilon_{F,0}=0.04$\ rad\ s$^{-1}$, (b): $f_{\mathbf{U}}=0.01 F_0$, $\varepsilon_{F,0}=0.16$\ rad\ s$^{-1}$, (c): $f_{\mathbf{U}}=0.01 F_0$, $\varepsilon_{F,0}=0.64$\ rad\ s$^{-1}$, (d): $f_{\mathbf{U}}=0.05 F_0$, $\varepsilon_{F,0}=0.04$\ rad\ s$^{-1}$, (e): $f_{\mathbf{U}}=0.05 F_0$, $\varepsilon_{F,0}=0.16$\ rad\ s$^{-1}$, (f): $f_{\mathbf{U}}=0.05 F_0$, $\varepsilon_{F,0}=0.64$\ rad\ s$^{-1}$.}
    \label{fig: jelly tracer two clusters}
\end{figure}

The ability of the swarm to cross the flow barrier is found to depend mainly on {the angular coupling constant} $\varepsilon_{F,0}$ and {velocity decoupling parameter} $f_{\mathbf{U}}$. {On the contrary, crossing performance }is largely independent of $f_F$ (see Fig.~\ref{fig: jelly tracer spread fraction}). In particular, we see that for $f_{\mathbf{U}}=0.05 F_0$ ($86\%$ weakening of counter current coupling at $F_j=0.3 F_0$) more jellyfish travel to the left gyre (Fig.~\ref{fig: jelly tracer spread fraction} panel (a)) than for $f_{\mathbf{U}}=0.01 F_0$ ($97\%$ weakening of counter current coupling at $F_j=0.3 F_0$). Interestingly, this difference occurs due to the emergence of secondary clusters in the right part of the tank (see for instance panels (b,c,f) in Fig.~\ref{fig: jelly tracer two clusters}). These clusters emerge due to a long-time equilibrium of orientations against the flow ($\delta_{\mathbf{U}}$) and towards prey ($\delta_F$). The upper cluster is less stable than the lower one because jellyfish in the upper right part of the flow are the last ones to become activated. As a consequence, they remain too slow for the flow in that region and experience a slow downstream transport. Ultimately, they come close to the upper central region of the tank where they feel the prey and cross the barrier. 

On the contrary, agents of the lower cluster experience a more diffuse flow and thus, are able to counteract advection. Ultimately they are able to approach the flow barrier where an increase in activity sparks a shift in the coupling parameters such that they cross over as well. However, those jellyfish also experience a small leakage of prey from left to right domain, which is transported by the lower boundary current in which they swim. This, on the one hand causes stronger bell strokes. On the other hand, the jellyfish start to orient partly towards the ambient prey. But, since the prey gradient and the direction of the nutritious flow are mostly not aligned, many jellyfish get dragged away. Some of them find a new equilibrium position further away from the flow barrier while others get transported with the current to the upper part of the barrier where they cross. Overall, some of the jellyfish remain inside the nutritious current, mainly on the lower right side of the tank. 

In fact, {coupling} values $\varepsilon_{F,0}>0.16$\ rad\ s$^{-1}$ make the escape from the lower secondary cluster more difficult while at the same time causing the swarm to have a small spread in the left domain (Fig.~\ref{fig: jelly tracer spread fraction} panel (b)). On the contrary, for $\varepsilon_{F,0}<0.16$\ rad\ s$^{-1}$, the jellyfish do not orient sufficiently towards prey and get trapped in the corners of the tank. This leaves us with a narrow parameter region of optimal $\varepsilon_{F,0}$ (grey regions in Fig.~\ref{fig: jelly tracer spread fraction}) out of which we choose $\varepsilon_{F,0}=0.16$\ rad\ s$^{-1}$.

\begin{figure}[!h!]
    \centering
    \includegraphics[width=\columnwidth]{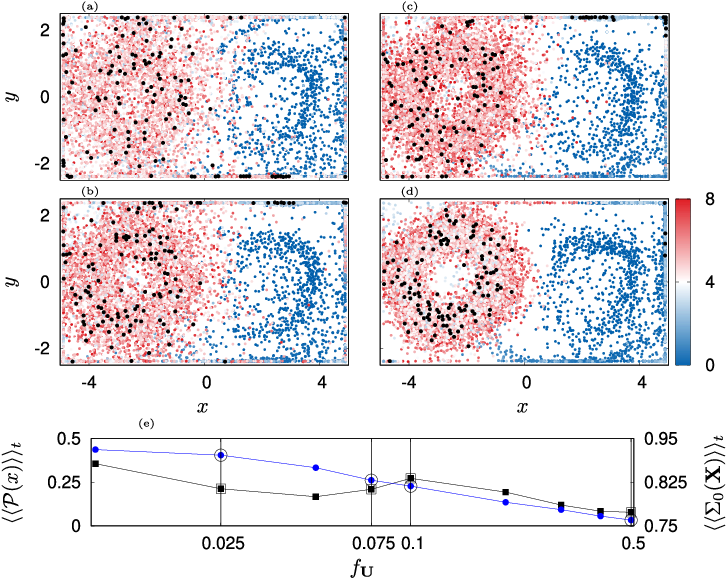}
    \caption{Depicted are scatter plots of jellyfish positions over $900$\ s and with an increment of $13$\ s for (a): $f_{\mathbf{U}}=0.025 F_0$, (b): $f_{\mathbf{U}}=0.075 F_0$, (c): $f_{\mathbf{U}}=0.1 F_0$, (d): $f_{\mathbf{U}}=0.5 F_0$. Other parameters are fixed according to optimal values given in Tab.~\ref{tab: params inputs}. Color indicates the activity ${\cal A}$. Black dots indicate the swarm in the end of the simulation. Panel (e) depicts the time averages of the agents fraction $\langle \langle {\cal P}(x) \rangle \rangle_t$ at $x>0$ (black squares, left ordinate) and of the spreading $\langle \langle \Sigma_0(\mathbf{X}) \rangle \rangle_t$ when $x<0$ (blue dots, right ordinate).}
    \label{fig: scan fa}
\end{figure}

Finally, we investigate {how the response to prey depends on the velocity decoupling parameter} $f_{\mathbf{U}}$. For this we consider just $\varepsilon_{F,0}=0.16$\ rad\ s$^{-1}$. Here, we find that because the value of $f_{\mathbf{U}}$ determines the instantaneous coupling strength $\varepsilon_{\mathbf{U}}(F)$, it simultaneously influences the fraction $\langle {\cal P} \rangle$ of remaining agents in the right gyre and the swarm spreading in the left gyre (Fig.~\ref{fig: scan fa} panels (a-d)). We see that for largely persistent counter-current swimming almost all agents cross the flow barrier ($f_{\mathbf{U}}=0.5 F_0$, $38\%$ weakening of counter current coupling at {prey concentration} $F_j=0.3 F_0$, Fig.~\ref{fig: scan fa} panel (d)). However they do not spread in the left domain but form a ring pattern. On the contrary, for earlier loss of counter-current orientation ($f_{\mathbf{U}}=[0.025, 0.075, 0.1] F_0$, $[92,80,75]\%$ weakening of counter current coupling at $F_j=0.3 F_0$, Fig.~\ref{fig: scan fa} panel (a,b,c)), secondary clusters remain in the right domain (Fig.~\ref{fig: scan fa} panels (a-c)) but the ring patter in the left gyre is less pronounced. Overall, this leads us to the conclusion that we consider $f_{\mathbf{U}}=0.05 F_0$ as an optimal parameter as it allows almost complete crossing of the swarm and reasonable spreading of the swarm in the left domain of the tank (see Fig.~\ref{fig: tracer optimal search}).

\section*{Conclusion}\label{sec: discussion}

In this paper we have proposed a paradigmatic model for jellyfish swarming based on active Brownian particles. The model can be readily incorporated in large-scale simulation models of the ocean and it provides a measurement paradigm to understand mechanisms of large scale structure formation based on agent responses, agent physiology and local interaction mechanisms. In its present form, Eqs.~(\ref{eq activity state}-\ref{eq: orientation state}) and Eqs.~(\ref{eq: responses eps}-\ref{eq attr rep}), the model is already able to reproduce paradigmatic jellyfish behavior reported in the literature. In particular, we have been able to simulate the counter-current swimming of \textit{R.~nomadica} \cite{malul2019levantine}. Using the kinematic tracer of absolute vorticity, we suggested a mechanism for the formation of long-lasting swarm patterns. Finally, we simulated the qualitatively reported behavior of jellyfish in a tank when prey is present \cite{arai1991attraction, hays2012high, albert2011s}. 

Here, we will discuss several aspects of swarming that have to be investigated in a future model. We largely dismissed jellyfish physiology of swimming and agent-agent interactions. The physiology of a jellyfish comprises three aspects. First, we expect that major parts of communication among jellyfish can be attributed to the turbulent vortex rings which are send into the water with a certain angular dependence $Q_V(\mathbf{x}_k,\mathbf{x}_j, \theta_k,\theta_j)$ (red arrow in Fig.~\ref{fig schematic model} panel (a)). 

Second, jellyfish possesses specific sensing capabilities. It has been reported that jellyfish occasionally bump into each other ''head on'' while at the same time being able to sense already faint pressure signals at a distance \cite{zafrirpersonal}. This suggests that jellyfish perceive their surrounding depending on the distribution of rophalia on the surface of the bell, introducing an angular dependence of sensitivity to perturbations. In the given framework, we suggest to introduce a function $Q_S(\mathbf{x}_k,\mathbf{x}_j, \theta_k,\theta_j)$ that corresponds to this distribution. Accordingly, the resulting physiological effect on agent-agent coupling, can be incorporated by a product function $Q(\mathbf{x}_k,\mathbf{x}_j, \theta_k,\theta_j)=Q_S(\mathbf{x}_k,\mathbf{x}_j, \theta_k,\theta_j) \cdot$ $Q_V(\mathbf{x}_k,\mathbf{x}_j, \theta_k,\theta_j)$. The function $Q(.)$ then mimics the previously envisaged angular dependence of coupling in attraction and repulsion Eq.~\eqref{eq attr rep} and in the functions $H_j(.)$ and ${\cal G}_j(.)$, Eq.~\eqref{eq swarming jelly}. 

Third, one needs to specify how the dynamics of the internal neuronal network reacts to external stimuli. This effect can be taken into account by the phase coupling function $H_j(.)$ of a medusa. It describes the response of the network phase, $\varphi$, Eq.~\eqref{eq swarming jelly} to external perturbations and has found widespread application in research on biological oscillators \cite{pikovsky2001synchronization,khalsa2003phase, brown2004phase, smeal2010phase, Kralemann_etal-13, rosenblum2018inferring, rosenblum2019nonlinear}. As a consequence of the directional dependence of agent-agent coupling, given by $Q(.)$, the phase response and the sensitivity response have to be disentangled simultaneously. Here, tracking techniques and phase-dynamics reconstruction have to be employed \cite{rife2001visual, smal2009quantitative, gengel2019phase}. 

Currently, we have considered only a simplified fluid-like soft-core repulsion that avoids any attractive effects. Here, further investigation of the radial dependence of the flow field around a medusa are needed. Beyond the angular dependence of communication, given by $Q(.)$, jellyfish might also be subject to inevitable effects of synchronization among each other if it fits their energy household \cite{ashraf2016synchronization, gruber2019bioluminescent}. This effect would involve the oscillatory phase of the bell, $\varphi$ and further modifies the attraction and repulsion in Eq.~\eqref{eq attr rep}. 

Generally, synchronization effects can be subtle because biological oscillators tend to only partially synchronize. From that perspective it is an intriguing open question to which extend synchronization of bell pulsation is present in jellyfish aggregations and whether it leads to a changes in swimming performance of jellyfish in blooms. For example, in our model the almost complete absence of a mechanism that could lead to synchrony of the bell oscillations causes significant performance losses of jellyfish clusters due to jamming. In fact, the only mechanisms through which the bell oscillations can couple in the model is present when agents search for prey. We consider this effect to be insignificant as it is caused by the interplay of prey concentration, repulsive interaction, activity dynamics and frequency response. On the contrary, bells stay effectively asynchronous such that densely packed agents push each other out of the way until they have sorted into groups where pushing becomes rare. This mechanism is know as social sorting \cite{helbing1995social}. Indeed, this effect might be partly visible where agents form highly dense clusters.

In our given model, walls and agents have the same effect on the positional dynamics, depending on the interaction distance $R_i$. We have seen during our numerical analysis that the choice of $R_i$ significantly influences the foraging of jellyfish in the secondary clusters {at the walls}. Thus, a further open question is the nature of the wall interactions for jellyfish.

We have completely ignored perturbations in the fluid due to the presence of agents. Our motivation here is that our model will serve as a parameterization of a sub-grid-scale process \cite{shchepetkin2005regional,adcroft2019gfdl, stegner2021cyclone}. As such, the local fluid perturbations of single jellyfish become irrelevant. Therefore, we take the local perturbations into account only by the angular noise $L_j$. One of the two essential parameters of this noise, $D_0$, was chosen and thus, introduced a time scale to which the coupling constants of orientation relate. To make the simulation more accurate, the actual statistical properties of the angular noise need to be investigated from passive observations. Furthermore, in a large-scale ocean simulation, the fluid-agent coupling will require a high-order interpolation scheme in time and space to provide accurate Lagrangian field values. In turn, the agent-fluid interaction shall rely on averaging over agents in a single cell.

The decision making of jellyfish in our model is realized by response functions ${\cal R}(.)$ that differ only by two parameters and are otherwise similar in shape (see panel (c) Fig.~\ref{fig: couplings}). We have employed this concept of response functions because of its great success in the description of cell membrane responses in computational neuroscience \cite{hodgkin1952quantitative, morris1981voltage}. In contrast to those systems where the membrane response was measurable directly, investigation of the response parameters will have to rely on indirect observations and data fitting methods such as machine learning \cite{kashinath2021physics}.

We have simulated a two-dimensional swarming dynamics. Thus, in its present form, the model Eqs.~(\ref{eq activity state}-\ref{eq: orientation state}) and Eqs.~(\ref{eq: responses eps}-\ref{eq attr rep}) can be used to capture the vertically averaged dynamics of a swarm and the dynamics of jellyfish in periods of reduced vertical migration. However, in reality, jellyfish swimming is influenced by the day night-cycle \cite{kaartvedt2007diel, mackie1981swimming}, its prey and environmental inputs such as temperature and salinity \cite{heim2019salinity} which cause vertical migration. Augmentation of the model dynamics to three dimensions is relatively straight forward and requires a second directional angle.

{We have chosen an over-damped positional dynamics Eq.~\eqref{eq: pos state} to model the swimming. By this we have in mind, that on larger scale, in the sea, the short time and length scales associated with inertia are not relevant. However, this assumption gives rise to two effects in our work. First, we observe regaining of mobility once the density of individuals becomes large enough (see Fig.~\ref{fig: passive agents phase transition}). Second, the phase in the velocity oscillation function $\beta(\varphi)$ Eq.~\eqref{eq: velocity} has only minimal effect on the positional dynamics as the time-average of velocity is completely determined by the shape parameter $J$. A future model might consider also inertia for jellyfish, including their velocity dynamics \cite{romanczuk2012active,tonertu}. However, such a model requires twice as much initial conditions (for the positions $\mathbf{x}_j(0)$ and velocities $\mathbf{v}_j(0)$), imposing even greater demands on any data-driven set of initial conditions.}

The turbulence avoidance and the prey searching represent a type of first-passage problem in which the supremum of the first passage times of single agents could be used to define a first passage time for the swarm to enter the target region \cite{redner2001guide}. However, we would like to point out that agent trajectories are not independent of each other and that the boundary of the target region can be expected to have a fractal, time-dependent structure according to the Reynolds number. 

Finally, we have investigated just a finite time horizon of maximal $15$ minutes due to numerical constraints. We think that in future analysis longer observation times are needed to capture long-term effects of swarming.

\section{Acknowledgements}

Erik Gengel is supported by a Minerva Fellowship of the Minerva Stiftung Gesellschaft fuer die Forschung mbH and thanks the postdoctoral exchange programme of the Universities Tel Aviv and Potsdam for financial support. Erik Gengel thanks Rok Cestnik, Vicky Verma, Subhajit Kar and Marcus Dahlenburg for helpful discussions. This article is thankfully dedicated to Prof.\ Dr.\ Arkady Pikovsky. Computations have been supported by the Israel Data Science Initiative.


\begin{thebibliography}{10}
\bibitem{van2014origin}
H.~Van~Iten, A.~C. Marques, J.~D.~M. Leme, M.~L.~F. Pacheco, M.~G. Simoes,
  Origin and early diversification of the \textit{Phylum Cnidaria Verrill}: major developments in the analysis of the taxon's proterozoic--cambrian history, Palaeontology 57~(4) (2014) 677--690.

\bibitem{helm2018evolution}
R.~R. Helm, Evolution and development of scyphozoan jellyfish, Biological
  Reviews 93~(2) (2018) 1228--1250.

\bibitem{lotan1992life}
A.~Lotan, R.~Ben-Hillel, Y.~Loya, Life cycle of \textit{Rhopilema nomadica}: a new immigrant scyphomedusan in the Mediterranean, Marine Biology 112~(2) (1992)
  237--242.

\bibitem{costello2021hydrodynamics}
J.~H. Costello, S.~P. Colin, J.~O. Dabiri, B.~J. Gemmell, K.~N. Lucas, K.~R.
  Sutherland, The hydrodynamics of jellyfish swimming, Annual Review of Marine
  Science 13 (2021) 375--396.

\bibitem{gemmell2013passive}
B.~J. Gemmell, J.~H. Costello, S.~P. Colin, C.~J. Stewart, J.~O. Dabiri,
  D.~Tafti, S.~Priya, Passive energy recapture in jellyfish contributes to
  propulsive advantage over other metazoans, Proceedings of the National
  Academy of Sciences 110~(44) (2013) 17904--17909.

\bibitem{angel2016local}
D.~L. Angel, D.~Edelist, S.~Freeman, Local perspectives on regional challenges:
  jellyfish proliferation and fish stock management along the Israeli
  Mediterranean coast, Regional Environmental Change 16~(2) (2016) 315--323.

\bibitem{schrope2012attack}
M.~Schrope, Attack of the blobs, Nature 482~(7383) (2012) 20.

\bibitem{streftaris2006alien}
N.~Streftaris, A.~Zenetos, Alien marine species in the Mediterranean-the 100
  ‘worst invasives’ and their impact, Mediterranean Marine Science 7~(1)
  (2006) 87--118.

\bibitem{nakar2011economic}
N.~Nakar, D.~Disegni, D.~Angel, Economic evaluation of jellyfish effects on the
  fishery sector—case study from the eastern Mediterranean, in: Proceedings
  of the Thirteenth Annual BIOECON Conference, Vol.~10, 2011, pp. 11--13.

\bibitem{attrill2007climate}
M.~J. Attrill, J.~Wright, M.~Edwards, Climate-related increases in jellyfish
  frequency suggest a more gelatinous future for the North Sea, Limnology and
  Oceanography 52~(1) (2007) 480--485.

\bibitem{zhang2012associations}
F.~Zhang, S.~Sun, X.~Jin, C.~Li, Associations of large jellyfish distributions
  with temperature and salinity in the Yellow Sea and East China Sea, in:
  Jellyfish Blooms IV, Springer, 2012, pp. 81--96.

\bibitem{baliarsingh2020review}
S.~K. Baliarsingh, A.~A. Lotliker, S.~Srichandan, A.~Samanta, N.~Kumar,
  T.~Nair, A review of jellyfish aggregations, focusing on India’s coastal
  waters, Ecological Processes 9~(1) (2020) 1--9.

\bibitem{heim2019salinity}
H.~Heim-Ballew, Z.~Olsen, Salinity and temperature influence on scyphozoan
  jellyfish abundance in the western Gulf of Mexico, Hydrobiologia 827~(1)
  (2019) 247--262.

\bibitem{edelist2020phenological}
D.~Edelist, T.~Guy-Haim, Z.~Kuplik, N.~Zuckerman, P.~Nemoy, D.~L. Angel,
  Phenological shift in swarming patterns of \textit{Rhopilema nomadica} in the eastern
  Mediterranean Sea, Journal of Plankton Research 42~(2) (2020) 211--219.

\bibitem{houghton2006developing}
J.~D. Houghton, T.~K. Doyle, J.~Davenport, G.~C. Hays, Developing a simple,
  rapid method for identifying and monitoring jellyfish aggregations from the
  air, Marine Ecology Progress Series 314 (2006) 159--170.

\bibitem{cimino2018jellyfish}
M.~A. Cimino, S.~Patris, G.~Ucharm, L.~J. Bell, E.~Terrill, Jellyfish
  distribution and abundance in relation to the physical habitat of Jellyfish
  Lake, Palau, Journal of Tropical Ecology 34~(1) (2018) 17--31.

\bibitem{brown2002forecasting}
C.~W. Brown, R.~R. Hood, Z.~Li, M.~B. Decker, T.~F. Gross, J.~E. Purcell, H.~V.
  Wang, Forecasting system predicts presence of sea nettles in Chesapeake Bay,
  Eos, Transactions American Geophysical Union 83~(30) (2002) 321--326.

\bibitem{ruiz2012model}
J.~Ruiz, L.~Prieto, D.~Astorga, A model for temperature control of jellyfish
  (cotylorhiza tuberculata) outbreaks: A causal analysis in a Mediterranean
  coastal lagoon, Ecological Modelling 233 (2012) 59--69.

\bibitem{fossette2015current}
S.~Fossette, A.~C. Gleiss, J.~Chalumeau, T.~Bastian, C.~D. Armstrong,
  S.~Vandenabeele, M.~Karpytchev, G.~C. Hays, Current-oriented swimming by
  jellyfish and its role in bloom maintenance, Current Biology 25~(3) (2015)
  342--347.

\bibitem{aouititen2019predicting}
M.~Aouititen, R.~Bekkali, D.~Nachit, X.~Luan, M.~Mrhraoui, Predicting jellyfish
  strandings in the Moroccan north-west Mediterranean coastline, European
  Scientific Journal 15~(2) (2019) 72--84.

\bibitem{prieto2015portuguese}
L.~Prieto, D.~Mac{\'\i}as, A.~Peliz, J.~Ruiz, Portuguese Man-of-War (physalia
  physalis) in the Mediterranean: A permanent invasion or a casual appearance?,
  Scientific reports 5~(1) (2015) 1--7.

\bibitem{nordstrom2019tracking}
B.~Nordstrom, M.~C. James, K.~Martin, B.~Worm, Tracking jellyfish and
  leatherback sea turtle seasonality through citizen science observers, Marine
  Ecology Progress Series 620 (2019) 15--32.

\bibitem{malul2019levantine}
D.~Malul, T.~Lotan, Y.~Makovsky, R.~Holzman, U.~Shavit, The Levantine jellyfish
  \textit{Rhopilema nomadica} and \textit{Rhizostoma pulmo} swim faster against the flow than with the flow, Scientific reports 9~(1) (2019) 1--6.

\bibitem{dabiri2005flow}
J.~O. Dabiri, S.~P. Colin, J.~H. Costello, M.~Gharib, Flow patterns generated
  by oblate medusan jellyfish: field measurements and laboratory analyses,
  Journal of Experimental Biology 208~(7) (2005) 1257--1265.

\bibitem{gemmell2015control}
B.~J. Gemmell, D.~R. Troolin, J.~H. Costello, S.~P. Colin, R.~A. Satterlie,
  Control of vortex rings for manoeuvrability, Journal of The Royal Society
  Interface 12~(108) (2015) 20150389.

\bibitem{hoover2017quantifying}
A.~P. Hoover, B.~E. Griffith, L.~A. Miller, Quantifying performance in the
  medusan mechanospace with an actively swimming three-dimensional jellyfish
  model, Journal of Fluid Mechanics 813 (2017) 1112--1155.

\bibitem{yuan2014numerical}
H.-Z. Yuan, S.~Shu, X.-D. Niu, M.~Li, Y.~Hu, A numerical study of jet
  propulsion of an oblate jellyfish using a momentum exchange-based immersed
  boundary-lattice Boltzmann method, Advances in Applied Mathematics and
  Mechanics 6~(3) (2014) 307--326.

\bibitem{wilson2009lagrangian}
M.~M. Wilson, J.~Peng, J.~O. Dabiri, J.~D. Eldredge, Lagrangian coherent
  structures in low Reynolds number swimming, Journal of Physics: Condensed
  Matter 21~(20) (2009) 204105.

\bibitem{hoover2019pump}
A.~P. Hoover, A.~J. Porras, L.~A. Miller, Pump or coast: the role of resonance
  and passive energy recapture in medusan swimming performance, Journal of
  Fluid Mechanics 863 (2019) 1031--1061.

\bibitem{dular2009numerical}
M.~Dular, T.~Bajcar, B.~{\v{S}}irok, Numerical investigation of flow in the
  vicinity of a swimming jellyfish, Engineering Applications of Computational
  Fluid Mechanics 3~(2) (2009) 258--270.

\bibitem{sahin2009arbitrary}
M.~Sahin, K.~Mohseni, An arbitrary Lagrangian--Eulerian formulation for the
  numerical simulation of flow patterns generated by the hydromedusa \textit{Aequorea
  victoria}, Journal of Computational Physics 228~(12) (2009) 4588--4605.

\bibitem{hoover2015numerical}
A.~Hoover, L.~Miller, A numerical study of the benefits of driving jellyfish
  bells at their natural frequency, Journal of theoretical biology 374 (2015)
  13--25.

\bibitem{park2015dynamics}
S.~G. Park, B.~Kim, J.~Lee, W.-X. Huang, H.~J. Sung, Dynamics of prolate
  jellyfish with a jet-based locomotion, Journal of Fluids and Structures 57
  (2015) 331--343.

\bibitem{hamner2009review}
W.~M. Hamner, M.~N. Dawson, A review and synthesis on the systematics and
  evolution of jellyfish blooms: advantageous aggregations and adaptive
  assemblages, Hydrobiologia 616~(1) (2009) 161--191.

\bibitem{arai1991attraction}
M.~N. Arai, Attraction of \textit{aurelia} and \textit{aequorea} to prey, in: Hydrobiologia, Vol. 216, Springer, 1991, pp. 363--366.

\bibitem{matanoski2001characterizing}
J.~Matanoski, R.~Hood, J.~Purcell, Characterizing the effect of prey on swimming and feeding efficiency of the scyphomedusa \textit{Chrysaora quinquecirrha}, Marine Biology 139~(1) (2001) 191--200.

\bibitem{omori2004taxonomic}
M.~Omori, M.~Kitamura, Taxonomic review of three Japanese species of edible
  jellyfish (scyphozoa: \textit{Rhizostomeae}), Plankton Biology and Ecology 51~(1)
  (2004) 36--51.

\bibitem{nooteboom2020resolution}
P.~D. Nooteboom, P.~Delandmeter, E.~van Sebille, P.~K. Bijl, H.~A. Dijkstra,
  A.~S. von~der Heydt, Resolution dependency of sinking Lagrangian particles in
  ocean general circulation models, PloS one 15~(9) (2020) e0238650.

\bibitem{bryan1995midlatitude}
F.~O. Bryan, C.~W. B{\"o}ning, W.~R. Holland, On the midlatitude circulation in
  a high-resolution model of the North Atlantic, Journal of physical
  oceanography 25~(3) (1995) 289--305.

\bibitem{delworth2012simulated}
T.~L. Delworth, A.~Rosati, W.~Anderson, A.~J. Adcroft, V.~Balaji, R.~Benson,
  K.~Dixon, S.~M. Griffies, H.-C. Lee, R.~C. Pacanowski, et~al., Simulated
  climate and climate change in the GFDL CM2.5 high-resolution coupled climate
  model, Journal of Climate 25~(8) (2012) 2755--2781.

\bibitem{schweitzer2003brownian}
F.~Schweitzer, J.~D. Farmer, Brownian agents and active particles: collective
  dynamics in the natural and social sciences, Vol.~1, Springer, 2003.

\bibitem{romanczuk2012active}
P.~Romanczuk, M.~B{\"a}r, W.~Ebeling, B.~Lindner, L.~Schimansky-Geier, Active
  Brownian particles, The European Physical Journal Special Topics 202~(1)
  (2012) 1--162.

\bibitem{o2017oscillators}
K.~P. O’Keeffe, H.~Hong, S.~H. Strogatz, Oscillators that sync and swarm,
  Nature communications 8~(1) (2017) 1--13.

\bibitem{o2019review}
K.~O'Keeffe, C.~Bettstetter, A review of swarmalators and their potential in
  bio-inspired computing, Micro-and Nanotechnology Sensors, Systems, and
  Applications XI 10982 (2019) 383--394.

\bibitem{hong2021coupling}
H.~Hong, K.~Yeo, H.~K. Lee, Coupling disorder in a population of swarmalators,
  Physical Review E 104~(4) (2021) 044214.

\bibitem{rodrigues2016kuramoto}
  F.~A. Rodrigues, T. Peron, P. Ji, J. Kurths, The Kuramoto model in complex networks, Physics Reports 610, 1--98, Elsevier, 2016

\bibitem{pikovsky2001synchronization}
A.~Pikovsky, M.~Rosenblum, J.~Kurths, Synchronization: a universal concept in
  nonlinear sciences, Cambridge University Press, 2001.

\bibitem{pikovsky2021transition}
A.~Pikovsky, Transition to synchrony in chiral active particles, Journal of
  Physics: Complexity 2~(2) (2021) 025009.

\bibitem{uriu2013dynamics}
K.~Uriu, S.~Ares, A.~C. Oates, L.~G. Morelli, Dynamics of mobile coupled phase
  oscillators, Physical Review E 87~(3) (2013) 032911.

\bibitem{zheng2021transition}
C.~Zheng, R.~Toenjes, A.~Pikovsky, Transition to synchrony in a
  three-dimensional swarming model with helical trajectories, Physical Review E
  104~(1) (2021) 014216.

\bibitem{zafrirpersonal}
Z.~Kuplik, D.~Angel, Personal communication.

\bibitem{rosenblum2001phase}
M.~Rosenblum, A.~Pikovsky, J.~Kurths, C.~Sch{\"a}fer, P.~A. Tass, Phase
  synchronization: from theory to data analysis, in: Handbook of biological
  physics, Vol.~4, Elsevier, 2001, pp. 279--321.

\bibitem{cestnikinfering2020}
R.~Cestnik, Inferring oscillatory dynamics from data, Ph.D. thesis, Vrije
  Universiteit Amsterdam (2020).

\bibitem{kralemann2007uncovering}
B.~Kralemann, L.~Cimponeriu, M.~Rosenblum, A.~Pikovsky, R.~Mrowka, Uncovering
  interaction of coupled oscillators from data, Phys. Rev. E 76~(5) (2007)
  055201.

\bibitem{gengel2021phase}
E.~Gengel, A.~Pikovsky, Phase reconstruction with iterated Hilbert transforms,
  in: Physics of Biological Oscillators, Springer, 2021, pp. 191--208.

\bibitem{smeal2010phase}
R.~M. Smeal, G.~B. Ermentrout, J.~A. White, Phase-response curves and
  synchronized neural networks, Philosophical Transactions of the Royal Society
  B: Biological Sciences 365~(1551) (2010) 2407--2422.

\bibitem{Rosenblum-Pikovsky-01}
M.~Rosenblum, A.~Pikovsky, Detecting direction of coupling in interacting
  oscillators, Phys. Rev. E 64~(4) (2001) 045202(R).

\bibitem{kashinath2021physics}
K.~Kashinath, M.~Mustafa, A.~Albert, J.~Wu, C.~Jiang, S.~Esmaeilzadeh,
  K.~Azizzadenesheli, R.~Wang, A.~Chattopadhyay, A.~Singh, et~al.,
  Physics-informed machine learning: case studies for weather and climate
  modelling, Philosophical Transactions of the Royal Society A 379~(2194)
  (2021) 20200093.

\bibitem{mcilwaine2021jellynet}
B.~Mcilwaine, M.~R. Casado, Jellynet: The convolutional neural network
  jellyfish bloom detector, International Journal of Applied Earth Observation
  and Geoinformation 97 (2021) 102279.

\bibitem{martin2020jellytoring}
M.~Martin-Abadal, A.~Ruiz-Frau, H.~Hinz, Y.~Gonzalez-Cid, Jellytoring:
  Real-time jellyfish monitoring based on deep learning object detection,
  Sensors 20~(6) (2020) 1708.

\bibitem{albajes2011jellyfish}
A.~Albajes-Eizagirre, L.~Romero, A.~Soria-Frisch, Q.~Vanhellemont, Jellyfish
  prediction of occurrence from remote sensing data and a non-linear pattern
  recognition approach, in: Remote Sensing for Agriculture, Ecosystems, and
  Hydrology XIII, Vol. 8174, SPIE, 2011, pp. 382--391.

\bibitem{albert2011s}
D.~J. Albert, What's on the mind of a jellyfish? a review of behavioural
  observations on \textit{Aurelia} sp. jellyfish, Neuroscience \& Biobehavioral Reviews
  35~(3) (2011) 474--482.

\bibitem{vicsek1995novel}
T.~Vicsek, A.~Czir{\'o}k, E.~Ben-Jacob, I.~Cohen, O.~Shochet, Novel type of
  phase transition in a system of self-driven particles, Physical review
  letters 75~(6) (1995) 1226.

\bibitem{wolgemuth2008collective}
C.~W. Wolgemuth, Collective swimming and the dynamics of bacterial turbulence,
  Biophysical journal 95~(4) (2008) 1564--1574.

\bibitem{aranson2013active}
I.~S. Aranson, Active colloids, Physics-Uspekhi 56~(1) (2013) 79.

\bibitem{cavagna2014bird}
A.~Cavagna, I.~Giardina, Bird flocks as condensed matter, Annu. Rev. Condens.
  Matter Phys. 5~(1) (2014) 183--207.

\bibitem{canizo2010collective}
J.~Canizo, J.~Carrillo, J.~Rosado, Collective behavior of animals: Swarming and
  complex patterns, Arbor 186~(1035-1049) (2010) 1.

\bibitem{ramaswamystatactive2010}
S.~Ramaswamy,
  The mechanics and statistics of active matter, Annual Review of Condensed Matter Physics 1~(1) (2010) 323--345.

\bibitem{stenhammar2017role}
J.~Stenhammar, C.~Nardini, R.~W. Nash, D.~Marenduzzo, A.~Morozov, Role of
  correlations in the collective behavior of microswimmer suspensions, Physical
  review letters 119~(2) (2017) 028005.

\bibitem{grossmann2014vortex}
R.~Gro{\ss}mann, P.~Romanczuk, M.~B{\"a}r, L.~Schimansky-Geier, Vortex arrays
  and mesoscale turbulence of self-propelled particles, Physical review letters
  113~(25) (2014) 258104.

\bibitem{reinken2019anisotropic}
H.~Reinken, S.~Heidenreich, M.~B{\"a}r, S.~H. Klapp, Anisotropic mesoscale
  turbulence and pattern formation in microswimmer suspensions induced by
  orienting external fields, New Journal of Physics 21~(1) (2019) 013037.

\bibitem{kampanis2006staggered}
N.~A. Kampanis, J.~A. Ekaterinaris, A staggered grid, high-order accurate
  method for the incompressible Navier--Stokes equations, Journal of
  Computational Physics 215~(2) (2006) 589--613.

\bibitem{kundufluiddyn}
D.~R.~D. Kundu K.~Pijush, Cohen M.~Ira, Fluid dynamics, Vol.~6 of Fluid
  dynamics, Elsevier, 2016.

\bibitem{okubo1971oceanic}
A.~Okubo, Oceanic diffusion diagrams, in: Deep sea research and oceanographic
  abstracts, Vol.~18, Elsevier, 1971, pp. 789--802.

\bibitem{mannella2002integration}
R.~Mannella, Integration of stochastic differential equations on a computer,
  International Journal of Modern Physics C 13~(09) (2002) 1177--1194.

\bibitem{press1992numerical}
W.~H. Press, S.~A. Teukolsky, W.~T. Vetterling, B.~P. Flannery, Numerical
  recipies in C, Vol.~3, Cambridge university press Cambridge, 1992.

\bibitem{adler2010quantifying}
J.~Adler, I.~Parmryd, Quantifying colocalization by correlation: the Pearson
  correlation coefficient is superior to the maMnder's overlap coefficient,
  Cytometry Part A 77~(8) (2010) 733--742.

\bibitem{mackie1981swimming}
G.~Mackie, R.~Larson, K.~Larson, L.~Passano, Swimming and vertical migration of \textit{Aurelia aurita} (l) in a deep tank, Marine \& Freshwater Behaviour \& Phy 7~(4) (1981) 321--329.

\bibitem{nath2017jellyfish}
R.~D. Nath, C.~N. Bedbrook, M.~J. Abrams, T.~Basinger, J.~S. Bois, D.~A.
  Prober, P.~W. Sternberg, V.~Gradinaru, L.~Goentoro, The jellyfish \textit{cassiopea}
  exhibits a sleep-like state, Current Biology 27~(19) (2017) 2984--2990.

\bibitem{ghosh2018kernel}
S.~Ghosh, Kernel smoothing: Principles, methods and applications, John Wiley \&
  Sons, 2018.

\bibitem{pallasdies2019single}
F.~Pallasdies, S.~Goedeke, W.~Braun, R.-M. Memmesheimer, From single neurons to
  behavior in the jellyfish \textit{Aurelia aurita}, Elife 8 (2019) e50084.

\bibitem{satterlie2011jellyfish}
R.~A. Satterlie, Do jellyfish have central nervous systems?, Journal of
  Experimental Biology 214~(8) (2011) 1215--1223.

\bibitem{garm2006rhopalia}
A.~Garm, P.~Ekstr{\"o}m, M.~Boudes, D.-E. Nilsson, Rhopalia are integrated
  parts of the central nervous system in box jellyfish, Cell and tissue
  research 325~(2) (2006) 333--343.

\bibitem{hoover2021neuromech}
A.~P. Hoover, N.~W. Xu, B.~J. Gemmell, S.~P. Colin, J.~H. Costello, J.~O.
  Dabiri, L.~A. Miller, Neuromechanical wave resonance in jellyfish swimming,
  Proceedings of the National Academy of Sciences 118~(11) (2021) e2020025118.

\bibitem{Winfree-80}
A.~T. Winfree, The Geometry of Biological Time, Springer, Berlin, 1980.

\bibitem{watanabe1994constants}
S.~Watanabe, S.~H. Strogatz, Constants of motion for superconducting Josephson
  arrays, Physica D: Nonlinear Phenomena 74~(3-4) (1994) 197--253.

\bibitem{kuramoto2003chemical}
Y.~Kuramoto, Chemical turbulence. Springer Berlin Heidelberg, 1984.

\bibitem{wilson2016isostable}
D.~Wilson, J.~Moehlis, Isostable reduction of periodic orbits, Physical Review
  E 94~(5) (2016) 052213.

\bibitem{hansel1993phase}
D.~Hansel, G.~Mato, C.~Meunier, Phase dynamics for weakly coupled
  Hodgkin-Huxley neurons, EPL (Europhysics Letters) 23~(5) (1993) 367.

\bibitem{levnajic2010phase}
Z.~Levnaji{\'c}, A.~Pikovsky, Phase resetting of collective rhythm in ensembles
  of oscillators, Phys. Rev. E 82~(5) (2010) 056202.

\bibitem{hagos2019synchronization}
Z.~Hagos, T.~Stankovski, J.~Newman, T.~Pereira, P.~V. McClintock,
  A.~Stefanovska, Synchronization transitions caused by time-varying coupling
  functions, Philosophical Transactions of the Royal Society A 377~(2160)
  (2019) 20190275.

\bibitem{topccu2018disentangling}
{\c{C}}.~Top{\c{c}}u, M.~Fr{\"u}hwirth, M.~Moser, M.~Rosenblum, A.~Pikovsky,
  Disentangling respiratory sinus arrhythmia in heart rate variability records,
  Physiological measurement 39~(5) (2018) 054002.

\bibitem{Kralemann_etal-13}
B.~Kralemann, M.~Fr\"uhwirth, A.~Pikovsky, M.~Rosenblum, T.~Kenner,
  J.~Schaefer, M.~Moser, In vivo cardiac phase response curve
  elucidates human respiratory heart rate variability, Nature Communications 4
  (2013) 2418.

\bibitem{bailey1983laboratory}
K.~Bailey, R.~Batty, A laboratory study of predation by \textit{Aurelia aurita} on
  larval herring (\textit{Clupea harengus}): experimental observations compared with
  model predictions, Marine Biology 72~(3) (1983) 295--301.

\bibitem{hays2012high}
G.~C. Hays, T.~Bastian, T.~K. Doyle, S.~Fossette, A.~C. Gleiss, M.~B. Gravenor,
  V.~J. Hobson, N.~E. Humphries, M.~K. Lilley, N.~G. Pade, et~al., High
  activity and L{\'e}vy searches: jellyfish can search the water column like
  fish, Proceedings of the Royal Society B: Biological Sciences 279~(1728)
  (2012) 465--473.

\bibitem{hodgkin1952quantitative}
A.~L. Hodgkin, A.~F. Huxley, A quantitative description of membrane current and
  its application to conduction and excitation in nerve, The Journal of
  physiology 117~(4) (1952) 500.

\bibitem{morris1981voltage}
C.~Morris, H.~Lecar, Voltage oscillations in the barnacle giant muscle fiber,
  Biophysical journal 35~(1) (1981) 193--213.

\bibitem{mauroy2013isostables}
A.~Mauroy, I.~Mezi{\'c}, J.~Moehlis, Isostables, isochrons, and Koopman
  spectrum for the action--angle representation of stable fixed point dynamics,
  Physica D: Nonlinear Phenomena 261 (2013) 19--30.

\bibitem{gengel2020high}
E.~Gengel, E.~Teichmann, M.~Rosenblum, A.~Pikovsky, High-order phase reduction
  for coupled oscillators, Journal of Physics: Complexity 2~(1) (2020) 015005.

\bibitem{buaria2020vortex}
D.~Buaria, E.~Bodenschatz, A.~Pumir, Vortex stretching and enstrophy production
  in high Reynolds number turbulence, Physical Review Fluids 5~(10) (2020)
  104602.

\bibitem{el2020modelling}
J.~El~Rahi, M.~P. Weeber, G.~El~Serafy, Modelling the effect of behavior on the
  distribution of the jellyfish mauve stinger (\textit{Pelagia noctiluca}) in the
  Balearic Sea using an individual-based model, Ecological Modelling 433 (2020)
  109230.

\bibitem{tonertu}
J.~Toner, Y.~Tu, Long-range order in a two-dimensional dynamical xy model: how
  birds fly together, Physical review letters 75~(23) (1995) 4326.

\bibitem{tevrugtpredict23}
M.~Te~Vrugt, J.~Bickmann, R.~Wittkowski, How to derive a predictive field
  theory for active Brownian particles: a step-by-step tutorial, Journal of
  Physics: Condensed Matter (2023).

\bibitem{negro2022inertial}
G.~Negro, C.~B. Caporusso, P.~Digregorio, G.~Gonnella, A.~Lamura, A.~Suma,
  Hydrodynamic effects on the liquid-hexatic transition of active colloids, The
  European Physical Journal E 45~(9) (2022) 75.

\bibitem{januswalt08}
A.~Walther, A.~H. M{\"u}ller, Janus particles, Soft matter 4~(4) (2008)
  663--668.

\bibitem{childs2008stability}
L.~M. Childs, S.~H. Strogatz, Stability diagram for the forced Kuramoto model,
  Chaos: An Interdisciplinary Journal of Nonlinear Science 18~(4) (2008)
  043128.

\bibitem{petkoski2012kuramoto}
S.~Petkoski, A.~Stefanovska, Kuramoto model with time-varying parameters,
  Physical Review E 86~(4) (2012) 046212.

\bibitem{locustariel15}
G.~Ariel, A.~Ayali, Locust collective motion and its modeling, PLOS
  computational Biology 11~(12) (2015) e1004522.

\bibitem{fireermen91}
B.~Ermentrout, An adaptive model for synchrony in the firefly \textit{Pteroptyx
  malaccae}, Journal of Mathematical Biology 29~(6) (1991) 571--585.

\bibitem{helbing1995social}
D.~Helbing, P.~Molnar, Social force model for pedestrian dynamics, Physical
  review E 51~(5) (1995) 4282.

\bibitem{drescher2011fluid}
K.~Drescher, J.~Dunkel, L.~H. Cisneros, S.~Ganguly, R.~E. Goldstein, Fluid
  dynamics and noise in bacterial cell--cell and cell--surface scattering,
  Proceedings of the National Academy of Sciences 108~(27) (2011) 10940--10945.

\bibitem{titelman2006feeding}
J.~Titelman, L.~J. Hansson, Feeding rates of the jellyfish \textit{Aurelia aurita} on
  fish larvae, Marine Biology 149~(2) (2006) 297--306.

\bibitem{rosenblum2019nonlinear}
M.~Rosenblum, A.~Pikovsky, Nonlinear phase coupling functions: a numerical
  study, Philosophical Transactions of the Royal Society A 377~(2160) (2019)
  20190093.

\bibitem{karlik2011performance}
B.~Karlik, A.~V. Olgac, Performance analysis of various activation functions in
  generalized mlp architectures of neural networks, International Journal of
  Artificial Intelligence and Expert Systems 1~(4) (2011) 111--122.

\bibitem{hansson1995behavioural}
L.~J. Hansson, K.~Kultima, Behavioural response of the scyphozoan jellyfish
  <\textit{Aurelia aurita} (l.) upon contact with the predatory jellyfish \textit{Cyanea
  capillata} (l.), Marine \& Freshwater Behaviour \& Phy 26~(2-4) (1995)
  131--137.

\bibitem{fier2018langevin}
G.~Fier, D.~Hansmann, R.~C. Buceta, Langevin equations for the run-and-tumble
  of swimming bacteria, Soft Matter 14~(19) (2018) 3945--3954.

\bibitem{saragosti2012modeling}
J.~Saragosti, P.~Silberzan, A.~Buguin, Modeling  \textit{E.~coli} tumbles by rotational
  diffusion. implications for chemotaxis, PloS one 7~(4) (2012) e35412.

\bibitem{polin2009chlamydomonas}
M.~Polin, I.~Tuval, K.~Drescher, J.~P. Gollub, R.~E. Goldstein, \textit{Chlamydomonas}
  swims with two “gears” in a eukaryotic version of run-and-tumble
  locomotion, Science 325~(5939) (2009) 487--490.

\bibitem{uhlenbeck1930theory}
G.~E. Uhlenbeck, L.~S. Ornstein, On the theory of the Brownian motion, Physical
  review 36~(5) (1930) 823.

\bibitem{costello1995flow}
J.~H. Costello, S.~Colin, Flow and feeding by swimming scyphomedusae, Marine
  Biology 124~(3) (1995) 399--406.

\bibitem{pikovsky2015dynamics}
A.~Pikovsky, M.~Rosenblum, Dynamics of globally coupled oscillators: Progress
  and perspectives, Chaos: An Interdisciplinary Journal of Nonlinear Science
  25~(9) (2015) 097616.

\bibitem{StrogatzNLD}
S.~H. Strogatz, Nonlinear dynamics and chaos with application to physics,
  biology, chemistry and engineering, Vol.~2, Westview, 2015.

\bibitem{degond2022environment}
P.~Degond, A.~Manhart, S.~Merino-Aceituno, D.~Peurichard, L.~Sala, How
  environment affects active particle swarms: a case study, arXiv preprint
  arXiv:2206.00329 (2022).

\bibitem{ashraf2016synchronization}
I.~Ashraf, R.~Godoy-Diana, J.~Halloy, B.~Collignon, B.~Thiria, Synchronization
  and collective swimming patterns in fish (\textit{Hemigrammus bleheri}), Journal of
  the Royal Society Interface 13~(123) (2016) 20160734.

\bibitem{alben2013efficient}
S.~Alben, L.~Miller, J.~Peng, Efficient kinematics for jet-propelled swimming,
  Journal of Fluid Mechanics 733 (2013) 100--133.

\bibitem{miles2019don}
J.~G. Miles, N.~A. Battista, Don't be jelly: Exploring effective jellyfish
  locomotion, arXiv preprint arXiv:1904.09340 (2019).

\bibitem{park2014simulation}
S.~G. Park, C.~B. Chang, W.-X. Huang, H.~J. Sung, Simulation of swimming oblate
  jellyfish with a paddling-based locomotion, Journal of Fluid Mechanics 748
  (2014) 731--755.

\bibitem{weeks1971d}
J.~Weeks, D.~chandler and H.~C.~Andersen, J. Chem. Phys 54 (1971) 5237.

\bibitem{buttinoni2013dynamical}
I.~Buttinoni, J.~Bialk{\'e}, F.~K{\"u}mmel, H.~L{\"o}wen, C.~Bechinger,
  T.~Speck, Dynamical clustering and phase separation in suspensions of
  self-propelled colloidal particles, Physical review letters 110~(23) (2013)
  238301.

\bibitem{rex2007lane}
M.~Rex, H.~L{\"o}wen, Lane formation in oppositely charged colloids driven by
  an electric field: Chaining and two-dimensional crystallization, Physical
  review E 75~(5) (2007) 051402.

\bibitem{wysocki2014cooperative}
A.~Wysocki, R.~G. Winkler, G.~Gompper, Cooperative motion of active Brownian
  spheres in three-dimensional dense suspensions, EPL (Europhysics Letters)
  105~(4) (2014) 48004.

\bibitem{giacche2010hydrodynamic}
D.~Giacch{\'e}, T.~Ishikawa, Hydrodynamic interaction of two unsteady model
  microorganisms, Journal of theoretical biology 267~(2) (2010) 252--263.

\bibitem{hamner1994sun}
W.~Hamner, P.~Hamner, S.~Strand, Sun-compass migration by \textit{Aurelia aurita}
  (scyphozoa): population retention and reproduction in Saanich Inlet, British
  Columbia, Marine Biology 119~(3) (1994) 347--356.

\bibitem{ghia1982high}
U.~Ghia, K.~N. Ghia, C.~Shin, High-re solutions for incompressible flow using
  the Navier-Stokes equations and a multigrid method, Journal of computational
  physics 48~(3) (1982) 387--411.

\bibitem{strandburg1988two}
K.~J. Strandburg, Two-dimensional melting, Reviews of modern physics 60~(1)
  (1988) 161.

\bibitem{gasser2010melting}
U.~Gasser, C.~Eisenmann, G.~Maret, P.~Keim, Melting of crystals in two
  dimensions, ChemPhysChem 11~(5) (2010) 963--970.

\bibitem{schmidle2012phase}
H.~Schmidle, C.~K. Hall, O.~D. Velev, S.~H. Klapp, Phase diagram of
  two-dimensional systems of dipole-like colloids, Soft Matter 8~(5) (2012)
  1521--1531.

\bibitem{hesse2014self}
J.~Hesse, T.~Gross, Self-organized criticality as a fundamental property of
  neural systems, Frontiers in systems neuroscience 8 (2014) 166.

\bibitem{mora2011biological}
T.~Mora, W.~Bialek, Are biological systems poised at criticality?, Journal of
  Statistical Physics 144~(2) (2011) 268--302.

\bibitem{khalsa2003phase}
S.~B.~S. Khalsa, M.~E. Jewett, C.~Cajochen, C.~A. Czeisler, A phase response
  curve to single bright light pulses in human subjects, The Journal of
  Physiology 549~(3) (2003) 945--952.

\bibitem{brown2004phase}
E.~Brown, J.~Moehlis, P.~Holmes, On the phase reduction and response dynamics
  of neural oscillator populations, Neural Computation 16~(4) (2004) 673--715.

\bibitem{rosenblum2018inferring}
M.~Rosenblum, C.~Rok, Inferring the phase response curve from observation of a
  continuously perturbed oscillator, Scientific Reports 8~(1) (2018) 13606.

\bibitem{rife2001visual}
J.~Rife, S.~M. Rock, Visual tracking of jellyfish in situ, in: Proceedings 2001
  International Conference on Image Processing (Cat. No. 01CH37205), Vol.~1,
  IEEE, 2001, pp. 289--292.

\bibitem{smal2009quantitative}
I.~Smal, M.~Loog, W.~Niessen, E.~Meijering, Quantitative comparison of spot
  detection methods in fluorescence microscopy, IEEE transactions on medical
  imaging 29~(2) (2009) 282--301.

\bibitem{gengel2019phase}
E.~Gengel, A.~Pikovsky, Phase demodulation with iterative Hilbert transform
  embeddings, Signal Processing 165 (2019) 115--127.

\bibitem{gruber2019bioluminescent}
D.~F. Gruber, B.~T. Phillips, R.~O’Brien, V.~Boominathan, A.~Veeraraghavan,
  G.~Vasan, P.~O’Brien, V.~A. Pieribone, J.~S. Sparks, Bioluminescent flashes
  drive nighttime schooling behavior and synchronized swimming dynamics in
  flashlight fish, PLoS One 14~(8) (2019) e0219852.

\bibitem{shchepetkin2005regional}
A.~F. Shchepetkin, J.~C. McWilliams, The regional oceanic modeling system
  (ROMS): a split-explicit, free-surface, topography-following-coordinate
  oceanic model, Ocean modelling 9~(4) (2005) 347--404.

\bibitem{adcroft2019gfdl}
A.~Adcroft, W.~Anderson, V.~Balaji, C.~Blanton, M.~Bushuk, C.~O. Dufour, J.~P.
  Dunne, S.~M. Griffies, R.~Hallberg, M.~J. Harrison, et~al., The GFDL global
  ocean and sea ice model OM4.0: Model description and simulation features,
  Journal of Advances in Modeling Earth Systems 11~(10) (2019) 3167--3211.

\bibitem{stegner2021cyclone}
A.~Stegner, B.~Le~Vu, F.~Dumas, M.~A. Ghannami, A.~Nicolle, C.~Durand,
  Y.~Faugere, Cyclone-anticyclone asymmetry of eddy detection on gridded
  altimetry product in the Mediterranean sea, Journal of Geophysical Research:
  Oceans 126~(9) (2021) e2021JC017475.

\bibitem{kaartvedt2007diel}
S.~Kaartvedt, T.~A. Klevjer, T.~Torgersen, T.~A. S{\o}rnes, A.~R{\o}stad, Diel
  vertical migration of individual jellyfish (\textit{periphylla periphylla}), Limnology
  and Oceanography 52~(3) (2007) 975--983.

\bibitem{redner2001guide}
S.~Redner, A guide to first-passage processes, Cambridge university press,
  2001.

\end{thebibliography}
\end{document}